\documentclass[a4paper,usenatbib]{mnras}
\usepackage{newtxtext,newtxmath}
\usepackage{comment}
\usepackage[T1]{fontenc}
\usepackage{ae,aecompl}
\usepackage{here}
\usepackage[pdftex]{graphicx}
\usepackage{amsmath}	
\usepackage{amssymb}	
\usepackage{mathrsfs}	%

\newcommand{\cm}{\mathrm{cm}}

\newcommand{\g}{\mathrm{g}}

\newcommand{\yr}{\mathrm{year}}
\newcommand{\yrs}{\mathrm{years}}

\newcommand{\AU}{{\rm AU}}
\newcommand{\Mearth}{M_{\mathrm{\oplus}}}

\newcommand{\Mjupiter}{M_{\mathrm{J}}}

\newcommand{\Msun}{M_{\mathrm{\odot}}}

\newcommand{\Gconst}{G}

\newcommand{\deriv}[2]{\frac{\mathrm{d} #1}{\mathrm{d} #2}}

\newcommand{\dderiv}[2]{\frac{\mathrm{d^2} #1}{\mathrm{d} {#2}^2}}

\newcommand{\pder}[2]{\frac{\partial {#1}}{\partial {#2}}}

\newcommand{\statave}[1]{\left\langle #1 \right\rangle} 

\newcommand{\axi}{a}
\newcommand{\baxi}{b}
\newcommand{\mm}{n}
\newcommand{\ecc}{e}
\newcommand{\inc}{i}
\newcommand{\lan}{\Omega_{\rm loa}}
\newcommand{\lpc}{\varpi}
\newcommand{\tra}{\psi}
\newcommand{\eca}{E}

\newcommand{\refeqs}[1]{{\rm Eq.~(\ref{#1})}}
\newcommand{\refsec}[1]{{\rm Section~\ref{#1}}}
\newcommand{\refapp}[1]{{\rm Appendix~\ref{#1}}}
\newcommand{\reffig}[1]{{\rm Fig.~\ref{#1}}}
\newcommand{\reftab}[1]{{\rm Table~ \ref{#1}}}

\newcommand{\Mstar}{M_{ {\rm s} }}
\newcommand{\Mplanet}{M_{ {\rm p} }}

\newcommand{\Mplts}{m_{ {\rm pl} }}
\newcommand{\Mcap}{M_{ {\rm cap} }}
\newcommand{\Mcaptot}{M_{\rm cap}^{\rm total}}

\newcommand{\Mdotplanet}{\dot{M}_{ {\rm p} }}
\newcommand{\MdotplanetR}{\dot{M}_{ {\rm p,R} }}
\newcommand{\MdotplanetS}{\dot{M}_{ {\rm p,S} }}
\newcommand{\Mdotcap}{\dot{M}_{ {\rm cap} }}

\newcommand{\Rplanet}{R_{ {\rm p} }}

\newcommand{\Rplts}{R_{ {\rm pl} }}
\newcommand{\RHill}{R_{ {\rm H} }}

\newcommand{\hill}{h}

\newcommand{\rplanet}{r_{ {\rm p} }}
\newcommand{\rplts}{r_{ {\rm pl} }}
\newcommand{\rstpl}{r_{ {\rm pl,s} }}
\newcommand{\rplpl}{r_{ {\rm pl,p} }}

\newcommand{\axiplanet}{a_{ {\rm p} }}

\newcommand{\vplts}{v_{ {\rm pl} }}
\newcommand{\vgas}{v_{ {\rm gas} }}
\newcommand{\vplgs}{u}
\newcommand{\vesc}{v_{ {\rm esc} }}
\newcommand{\vrel}{v_{ {\rm rel} }}

\newcommand{\vKep}{v_{ {\rm K}} }
\newcommand{\TKep}{T_{ {\rm K}} }
\newcommand{\OmegaK}{\Omega_{\rm K}}
\newcommand{\Omegap}{\Omega_{\rm p}}

\newcommand{\Sigmagas}{\Sigma_{\rm gas}}
\newcommand{\Sigmazero}{\Sigma_{\rm 0}}
\newcommand{\Sigmaf}{f_{\rm disc}}
\newcommand{\Sigmadust}{\Sigma_{\rm pl}}

\newcommand{\fgas}{f_{\rm gas}}
\newcommand{\cd}{C_{ {\rm D} }}
\newcommand{\rhogas}{\rho_{ {\rm gas} }}
\newcommand{\rhodust}{\rho_{ {\rm pl} }}
\newcommand{\taudamp}{\tau_{{\rm damp}}}
\newcommand{\taudampaxi}{\tau_{{\rm damp,}a}}
\newcommand{\taudampecc}{\tau_{{\rm damp,}e}}
\newcommand{\taudampinc}{\tau_{{\rm damp,}i}}
\newcommand{\tauacc}{\tau_{{\rm grow}}}
\newcommand{\etadisk}{\eta_{{\rm gas}}}

\newcommand{\csound}{c_{ {\rm s} }}

\newcommand{\Pgas}{P_{{\rm gas}}}

\newcommand{\Reyn}{\mathcal{R}}
\newcommand{\Mach}{\mathcal{M}}
\newcommand{\lmfp}{l_{\rm p}}

\newcommand{\Ocrr}{\omega}

\newcommand{\alphadisk}{\alpha_{{\rm disk}} }
\newcommand{\betadisk}{\beta_{{\rm disk}} }

\newcommand{\Sigmath}{\Sigma_{\rm th}}
\newcommand{\Sigmagap}{\Sigma_{ {\rm gap} }}
\newcommand{\Sigmamin}{\Sigma_{ {\rm min} }}
\newcommand{\hscale}{h_{ {\rm g} }}
\newcommand{\alphaPPD}{\alpha_{ {\rm \nu} }}
\newcommand{\DeltaGap}{\Delta_{ {\rm gap} }}
\newcommand{\Kp}{K^{'}}
\newcommand{\K}{K}
\newcommand{\Rone}{\Delta R_1}
\newcommand{\Rtwo}{\Delta R_2}

\newcommand{\Deltat}{\Delta t}
\newcommand{\rhoave}{\rho_{{\rm ave}}}

\newcommand{\Mpint}{M_{{\rm p,int}}}
\newcommand{\Mpfnl}{M_{{\rm p,fnl}}}

\newcommand{\Mupd}{{\rm model~A}}
\newcommand{\Mcdt}{{\rm model~B}} 
\newcommand{\Mgwoe}{{\rm model~B1}}
\newcommand{\Mgwe}{{\rm model~B2}} 
\newcommand{\ModelC}{{\rm model~C}} 
\newcommand{\ModelCa}{{\rm model~C0}} 
\newcommand{\ModelCb}{{\rm model~C1}} 
\newcommand{\ModelCc}{{\rm model~C2}} 
\newcommand{\ModelCd}{{\rm model~C3}} 
\newcommand{\vgasm}{v_{\rm gas,m}}
\newcommand{\xm}{x_{\rm m}}

\newcommand{\nplts}{n_\mathrm{pl}}
\newcommand{\collsec}{\Gamma_{\rm col}}
\newcommand{\Rcap}{R_{\rm cap}}
\newcommand{\hscaleplts}{h_{\rm pl}}

\newcommand{\Sigmaeff}{\Sigma_{ {\rm eff} }}
\newcommand{\Ejacobi}{E_{ {\rm J} }}
\newcommand{\Ejacobinrm}{\tilde{E}_{ {\rm J} }}
\newcommand{\Ujacobi}{U_{ {\rm J} }}

\newcommand{\vpltsp}{v_{ {\rm pl} }^{\prime}}

\newcommand{\fGwoE}{f_{\rm enh,B1} } 
\newcommand{\fGwE}{f_{\rm enh,B2} } 
\newcommand{\McapUD}{M_{\rm A}^{\rm total} } 
\newcommand{\McapGwoE}{M_{\rm B1}^{\rm total} } 
\newcommand{\McapGwE}{M_{\rm B2}^{\rm total} } 

\newcommand{\xgap}{x_{{\rm gap}}}

\newcommand{\Ejdotnrm}{{\rm d} \Ejacobinrm / {\rm d} t}

\newcommand{\taudampzero}{\tau_{{\rm 0}}}

\newcommand{\bnrm}{\tilde{b}} 
\newcommand{\eccnrm}{\tilde{e}}
\newcommand{\incnrm}{\tilde{i}}

\newcommand{\Force}{F}
\newcommand{\ForceR}{F_{\rm R}}
\newcommand{\ForceP}{F_{\rm \psi}}
\newcommand{\Forcez}{F_{\rm \zeta}}
\newcommand{\vplR}{v_{\rm pl,R}}
\newcommand{\vplP}{v_{\rm pl,\psi}}
\newcommand{\vplz}{v_{\rm pl,\zeta}}
\newcommand{\vgasR}{v_{\rm gas,R}}
\newcommand{\vgasP}{v_{\rm gas,\psi}}
\newcommand{\vgasz}{v_{\rm gas,\zeta}}
\newcommand{\urelR}{u_{\rm R}}
\newcommand{\urelP}{u_{\rm \phi}}
\newcommand{\urelz}{u_{\rm \zeta}}

\newcommand{\rr}{l}
\newcommand{\angeps}{\varepsilon}
\newcommand{\angdel}{\delta}

\newcommand{\taupep}{\tau_{{\rm disp}}}
\newcommand{\tcont}{t_{{\rm c}}}

\newcommand{\Sigmazeroint}{\Sigma_{{\rm 0,int}}}

\newcommand{\etats}{\eta_{{\rm ts}} }
\newcommand{\dacc}{\dot{ {\bf a} }}
\newcommand{\ddacc}{ {\bf a}^{(2)} }
\newcommand{\dddacc}{ {\bf a}^{(3)} }
\newcommand{\EngErr}{ \Delta E_{{\rm err}} }

\newcommand{\fei}[2]{f_{\rm e,i} ( #1, #2 )}
\newcommand{\axiin}{a_{\rm in}}
\newcommand{\axiout}{a_{\rm out}}
\newcommand{\axiinzero}{a_{\rm in,0}}
\newcommand{\axioutzero}{a_{\rm out,0}}
\newcommand{\Mdust}{M_{\rm solid}}

\newcommand{\Nplts}{N_{\rm pl}}
\newcommand{\Ejacobizero}{E_{ {\rm Jacobi,0} }}

\newcommand{\revised}[1]{{#1}}

\usepackage{ulem}

\title[Capture of Solids by Growing Protoplanet]{Capture of Solids by Growing Proto-gas Giants 
: Effects of Gap Formation and Supply-limited Growth}

\author[S. Shibata and M. Ikoma]{
Sho Shibata,$^{1}$\thanks{E-mail: s.shibata@eps.s.u-tokyo.ac.jp}
Masahiro Ikoma$^{1, 2}$ \\
$^{1}$Department of Earth and Planetary Science, Graduate School of Science, The University of Tokyo, 7-3-1 Hongo, Bunkyo-ku, Tokyo 113-0033, Japan \\
$^{2}$Research Center for the Early Universe (RESCEU),
Graduate School of Science, The University of Tokyo, 7-3-1 Hongo, 
Bunkyo-ku, Tokyo 113-0033, Japan
}

\date{Accepted XXX. Received YYY; in original form ZZZ}

\pubyear{2019}

\begin{document}
\label{firstpage}
\pagerange{\pageref{firstpage}--\pageref{lastpage}}
\maketitle

\begin{abstract}
Studies of internal structure of gas giant planets suggest that their envelopes are enriched with heavier elements than hydrogen and helium relative to their central stars.
Such enrichment likely occurred by solid accretion during late formation stages of gas giant planets in which gas accretion dominates protoplanetary growth.
Some previous studies performed orbital integration of planetesimals around a growing protoplanet with the assumption of an uniform circumstellar disc to investigate how efficiently the protoplanet captures planetesimals.
However, not only planetesimals but also disc gas are gravitationally perturbed by the protoplanet in its late formation stages, resulting in gap opening in the circumstellar disc.
In this study, we investigate the effects of such gap formation on the capture of planetesimals by performing dynamical simulations of planetesimals around a growing proto-gas giant planet.
Gap formation reduces the surface density of disc gas, makes a steep pressure gradient and limits the growth rate of the protoplanet.
We find that the first effect enhances the capture of planetesimals, while the others reduce it.
Consequently the amount of planetesimals captured during the gas accretion is estimated to be at most $\sim~3~\Mearth$.
We conclude that the in-situ capture of planetesimals needs the initial solid surface density more than five times higher than that of the minimum mass solar nebula for explaining the inferred large amount of heavy element in Jupiter.
For highly dense warm Jupiters, we would need additional processes enhancing the capture and/or supply of planetesimals.

\end{abstract}




\section{Introduction}
\label{sec:Introduction}
For understanding of the formation of gas giant planets, planetary internal composition provides crucial information.
Many studies of the internal structure of Jupiter and Saturn indicate that the gas giant planets contain much more heavy elements (or ``metals'') in their interiors, relative to the solar composition \citep[][references therein]{Guillot+2014}. 
Indeed, using the updated gravitational moments observed by the {\it Juno} spacecraft, \citet{Wahl+2017} recently estimated that the total mass of heavy elements contained inside Jupiter is as much as 24--46~$\Mearth$, while Jupiter would contain only several $\Mearth$ if its bulk composition were the solar.
Saturn's heavy-element content is also estimated to be as large as 13--28~$\Mearth$ \citep[][]{Saumon+2004}.
Likewise, for exoplanets, internal structure modeling of warm Jupiters based on their mass-radius relationships suggests that they contain greater amounts of heavy elements ($\sim10$--100~$\Mearth$) than their host stars \citep[][]{Guillot+2006,Miller+2011,Thorngren+2016}.
Importantly, although the core-envelope partitioning of heavy elements is uncertain, not all the heavy elements are contained in the core, but the envelope is highly enriched with heavy elements in most of the internal structure models consistent with observations.

From a viewpoint of planetary accretion, it is also difficult to form cores massive enough to account for the inferred amounts of heavy elements ($\gtrsim 20 \Mearth$), meaning that large fractions of the heavy elements are contained in the envelopes. 
In the core accretion model for planet formation, once a solid core grows to a critical mass, runaway gas accretion occurs to form a massive envelope \citep[][]{Mizuno+1980,Bodenheimer+1986}. 
In such a dense envelope, planetesimals are vaporized before reaching the core \citep[][]{Podolak+1988,Pollack+1996} and the heavy elements are distributed through the envelope. 
Although the critical core mass increases with core accretion rate, formation of such massive cores requires unrealistically high rates of core accretion \citep[][]{Ikoma+2000,Ikoma+2006b}. 
Conversely, if envelopes are highly enriched with heavy elements, the critical core mass is greatly reduced \citep[][]{Hori+2011, Venturini+2015}.

Hence, the enrichment of heavy elements inside gas giant planets is considered to occur in gas accretion phases after the critical core mass is reached.
A protoplanet grows in mass via gas accretion and, thus, expands its feeding zone.
This results basically in further accretion of planetesimals, provided dynamical effects of the protoplanet on planetesimals are ignored \citep[][]{Pollack+1996}.
However, in such gas accretion stages, since the protoplanet's mass is usually comparable with the total mass of the surrounding planetesimals, the protoplanet has a non-negligible, dynamical effect on the planetesimals.

Dynamical capture process of planetesimals by a growing proto-gas giant planet was first studied by \citet{Zhou+2007}.
They performed numerical integration of orbital changes of planetesimals (massless test-particles) around a protoplanet growing at an assumed rate in a gaseous protoplanetary disc.
They demonstrated that planetesimals are stirred gravitationally by the protoplanet and, then, change their orbits (i.e., migrate) because of disc-gas drag. 
Consequently, many of the planetesimals go out of the protoplanet's feeding zone without being engulfed, ending up with a gap around the protoplanet.
Also, some of them exterior to the outer boundary of the feeding zone are trapped at mean motion resonances to the protoplanet.
Thus, the capture rate of planetesimals depends on the expansion of the protoplanet's feeding zone and the excitation and damping of velocity dispersion of planetesimals by gravitational stirring and gas drag, respectively.
\citet{Shiraishi+2008} conducted a systematic investigation of such capture processes by similar simulations to \citet{Zhou+2007} and gave detailed interpretation to the capture mechanisms.
In addition, they derived semi-analytical formulae for the capture rate of planetesimals, which reproduced their simulation results well (also see Section~\ref{sec:Discussion_SI}).

\revised{Such} a hypothesis of dynamical capture of planetesimals during the gas accretion phases may certainly explain the observed excess of heavy elements qualitatively, but not quantitatively.
According to the above two studies, the total mass of planetesimals that the protoplanet captures till growing to the Jupiter mass is as small as $\sim$4--8~$\Mearth$ even in the case of a relatively massive circumstellar disc being about three times heavier than the minimum-mass solar nebula \citep[\revised{MMSN},][]{Hayashi1981}. 
Such heavy element contents of the envelope are obviously lower than those inferred from recent internal structure modeling.
Also, regarding the metallicities of warm gas giants, \citet{Hasegawa+2018} recently showed that not their absolute values but the correlation with host stars' metallicities are consistent with theoretical prediction derived by use of the semi-analytical formulae from \citet{Shiraishi+2008}.

Those previous studies showed gas drag for damping the velocity dispersion of planetesimals has a crucial effect on the dynamical capture of planetesimals by a protoplanet. 
However, they simply assumed that the gaseous disc was uniform, namely, the gas surface density was spatially constant. 
In reality, protoplanets with mass of several tens of $\Mearth$ have gravitational effects not only on planetesimals, but also on disc gas.
One of the important effects is the formation of a gap near the protoplanet' orbit in the circumstellar gas disc \citep[][]{Lin+1979,Goldreich+1980}.
\revised{Since gap formation brings about changes in the efficiency of damping of planetesimals' random velocities by disc gas drag and in the rate of the protoplanet's growth, it would affect the capture of planetesimals}.
In some cases, the gap becomes wider than the feeding zone and the surface density in the feeding zone becomes lower by several orders of magnitude than in the unperturbed disc \citep{Kanagawa+2016,Kanagawa+2017}.
This is important partly because the thinner the disc gas is, the higher the capture rate of planetesimals \citep[see][]{Zhou+2007,Shiraishi+2008} and partly because the modified gas density profile affects the direction and velocity of radial drift of planetesimals \citep[e.g.][]{Adachi+1976}.
In addition, once a gap is opened in the circumstellar disc, the growth of the protoplanet slows down \citep{Tanigawa+2007,Tanigawa+2016}, which, conversely, reduces the capture rate of planetesimals \citep[][]{Shiraishi+2008}.
Thus, it is difficult to know in advance whether and how the capture rate of planetesimals increases or decreases by the effect of gap opening.

The main purpose of this study is to quantify the effects of the gap formation on the capture of planetesimals by a growing proto-gas giant planet.
In \refsec{sec:Methods}, we describe the basic model and settings used in this study.
\revised{
We investigate the capture of planetesimals step by step.
In \refsec{sec:Model1}, we consider a protoplanet growing at a constant rate in an unperturbed circumstellar disc. 
We review the analytical interpretation for the capture of planetesimals given by \citet{Shiraishi+2008}, because that is helpful for understanding our numerical results. Then, we compare results of our numerical integration with their analytical formulae for the capture rate of planetesimals.
}
After that, we consider the effects of the presence of a gap in the circumstellar disc in \refsec{sec:Model2} and, furthermore, add the effects of time-variant rates of protoplanet growth regulated by disc gas supply in \refsec{sec:Model3}.
In \refsec{sec:Discussion}, we make a brief discussion of other effects that we have ignored. 
Finally we summarise and conclude this study in \refsec{sec:Summary}.

\section{Basic Method and Model}
\label{sec:Methods}
\subsection{Model Settings}\label{sec:Method_BS}

In order to investigate the capture of planetesimals by a growing protoplanet, we perform orbital integration of a protoplanet of mass $\Mplanet$ and 10,000 planetesimals of mass $\Mplts$ around a central star of mass $\Mstar$. 
The protoplanet is assumed to grow only via gas accretion with a rate $\Mdotplanet$, which means that planetesimal accretion makes no contribution to protoplanetary growth. 
Planetesimals are treated as test particles, their mutual gravitational interaction being ignored, and are subject to gas drag in the circumstellar (or protoplanetary) disc.
The protoplanet and central star orbit their common centre of mass; their orbits are assumed to be circular.
Here we \revised{set} that $\Mstar$ = 1~$\Msun$ and \revised{the semi-major axis of the protoplanet $\axiplanet$} is 5.2~AU.

The equation of motion for a planetesimal is
\begin{equation}\label{eq:equation_of_motion}
   \dderiv{ {\bf \rplts} }{t} = -\Gconst \frac{\Mstar}{\rstpl^3} {\bf \rstpl} -\Gconst \frac{\Mplanet}{\rplpl^3} {\bf \rplpl} + {\bf \fgas},
\end{equation}
where 
${\bf \rplts}$, ${\bf \rstpl}$ and ${\bf \rplpl}$ are the position vectors of the planetesimal relative to the mass centre, relative to the star, and relative to the protoplanet, respectively, $r_k = |{\bf r}_k|$ ($k$ = "pl", "pl,s", and "pl,p"), $t$ is the time, $\Gconst$ is the gravitational constant, and $\bf \fgas$ is the gas drag given by 
\begin{equation}\label{eq:reduced_gas_drag}
  {\bf \fgas} = - \frac{1}{2 \Mplts} \cd \pi {\Rplts}^2 \rhogas \vplgs {\bf \vplgs}.
\end{equation}
Here ${\bf \vplgs}$ is the planetesimal velocity relative to the ambient gas, $\cd$ is the non-dimensional drag coefficient, $\rhogas$ is the gas density, $\Rplts$ is the planetesimal radius, and $\vplgs = |{\bf \vplgs}|$.
In general, $\cd$ is a function of the Reynolds number $\Reyn$ (= $2 \rhogas \Rplts \vplgs  / \mu$; $\mu$ being dynamic viscosity) and 
the Mach number $\Mach$ (= $\vplts / \csound$; $\csound$ being the sound speed).
The dynamic viscosity is given by $\mu = (1/3) \rhogas \csound \lmfp$, where $\lmfp$ is the mean free path of gas molecules for which we adopt the collision cross section of hydrogen molecules (= $2 \times 10^{-15} {\cm}^2$).
In this study, we use an approximated formulae for $\cd$ written as \citep[e.g.][]{Tanigawa+2014}
\begin{align}\label{eq:Drag_Coefficient_Tanigawa}
	\cd \simeq \left[ \left( \frac{24}{\Reyn} + \frac{40}{10+\Reyn} \right)^{-1} + \frac{3 \Mach}{8} \right]^{-1} + \frac{ (2-\Ocrr) \Mach}{1+\Mach} + \Ocrr,
\end{align}
where $\Ocrr$ is a correction factor, the value of which is 0.4 for $\Reyn < 2 \times 10^5$ and 0.2 for $\Reyn > 2 \times 10^5$. 

The circumstellar disc is assumed to be vertically isothermal and, thus, the gas density $\rhogas$ is expressed with the surface density $\Sigmagas$ and the scale height of disc gas $\hscale$ as a function of the height from the mid-plane $z$ as
\begin{equation}\label{eq:Method_volume}
   \rhogas = \frac{\Sigmagas}{\sqrt{2 \pi} \hscale} \exp \left( -\frac{z^2}{2 {\hscale}^2} \right).
\end{equation}
For convenience we define the radial dependence of surface density $\Sigmaf (\rplts)$ as 
\begin{align}\label{eq:Method_Surface_Density}
    \Sigmaf (\rplts) = \Sigmagas (\rplts) / \Sigmazero,   
\end{align}
where $\Sigmazero$ is a reference surface density.
The scale height is defined as
\begin{equation}\label{eq:def_scale_height}
	\hscale \equiv \frac{\csound}{\OmegaK},
\end{equation}
where $\OmegaK$ is the Keplerian angular velocity.

The gas in the circumstellar disc rotates with a velocity that differs from the Keplerian velocity $\vKep$, because of pressure gradient; namely
\begin{equation}\label{eq:sub-Kepler}
	\vgas = \vKep \left( 1-\etadisk \right)
\end{equation}
with $\etadisk$ defined as
\begin{eqnarray}
	\etadisk	&\equiv&	\mathbf{-} \frac{1}{2} \left( \frac{\hscale}{r} \right)^2 \deriv{ \ln \Pgas}{\ln r}, 
\end{eqnarray}
where $r$ is the radial distance from the central star and $\Pgas$ is the gas pressure.
For deriving the above equation, we have assumed $\etadisk \ll 1$ and used the ideal-gas relation $\csound^2$ = $\Pgas / \rhogas$.
Using this relation for $\csound$ and Eq.~(\ref{eq:Method_volume}) for $\Sigmagas$, $\etadisk$ can be written as
\begin{eqnarray}\label{eq:eta_disk}
	\etadisk	
	&=&		\frac{1}{2} \left( \frac{\hscale}{r} \right)^2 
	\left[ 
		\frac{3}{2} \left( 1 - \frac{z^2}{{\hscale}^2} \right) +
		\alphadisk + 
		\betadisk \left(1+ \frac{z^2}{{\hscale}^2}  \right) 
	\right],
\end{eqnarray}
where
\begin{equation}\label{eq:def_a_b}
	\alphadisk	\equiv - \deriv{ \ln \Sigmagas }{ \ln r }, 
	\,\,\,
	\betadisk	\equiv - \deriv{ \ln \csound }{ \ln r }.
\end{equation}
In all the simulations below, we set $\Sigmazero = 1.43~\times~10^2~\g~{\cm}^{-3}$ and $\csound = 6.52~\times~10^4~\cm~{\rm s}^{-1}$, which correspond to the values at 5.2~$\AU$ in MMSN \citep{Hayashi1981}.
Then, the aspect ratio of the circumstellar gas disc at the protoplanet orbit comes out to be $\hscale / \rplanet \simeq 0.05$.

\subsection{Numerical integration of planetesimal orbits}
We integrate the equation of motion, \refeqs{eq:equation_of_motion}, for 10 000 ($\equiv \Nplts$) planetesimals numerically, using the forth-order-Hermite integration scheme \citep{Makino+1992}.
For timesteps $\Delta t$, we adopt the method of \citet{Aarseth1985b}, namely,
\begin{equation}\label{eq:Timestep}
	\Deltat = \etats \sqrt{ \frac{| {\bf a} | | \ddacc | + | \dacc |^2}{| \dacc | | \dddacc | + | \ddacc |^2} },
\end{equation}
where ${\bf a}$ is the acceleration of the planetesimal (= $\ddot{\bf r}_\mathrm{pl}$), ${\bf a}^{(k)}$ is the $k$th derivative of ${\bf a}$, and $\etats$ is an accuracy-controlling parameter. 
We have performed a benchmark test of the numerical code that we have newly developed in this study (see \refapp{App_BMT}).
In the benchmark test, we have confirmed that $\etats = 0.01$ yields sufficiently small energy error and, thus, use the value in all the numerical calculations presented below.

We judge that the planetesimal is trapped in the protoplanet, once (i) the planetesimal enters the protoplanet's envelope or (ii) the Jacobi energy, $\Ejacobi$, is negative in the Hill sphere (see \refsec{sec:Orb} for the definition of $\Ejacobi$), and, then, stop the orbital integration for the planetesimal.
We set the physical radius of the protoplanet $\Rplanet$ as 
\begin{align}\label{eq:Method_Physical_Radius}
    \Rplanet = \left(\frac{3 \Mplanet}{4 \pi \rhoave}\right)^{1/3},
\end{align}
where $\rhoave$ is the protoplanet's mean density.
In our simulations, we set $\rhoave = 0.125~\g \, \cm^{-3}$, which gives twice Jupiter's radius for Jupiter's mass.
We will discuss the validity of this setting later.

\subsection{Initial distribution of planetesimals}
\begin{figure}
    \begin{center}
    \includegraphics[width=80mm]{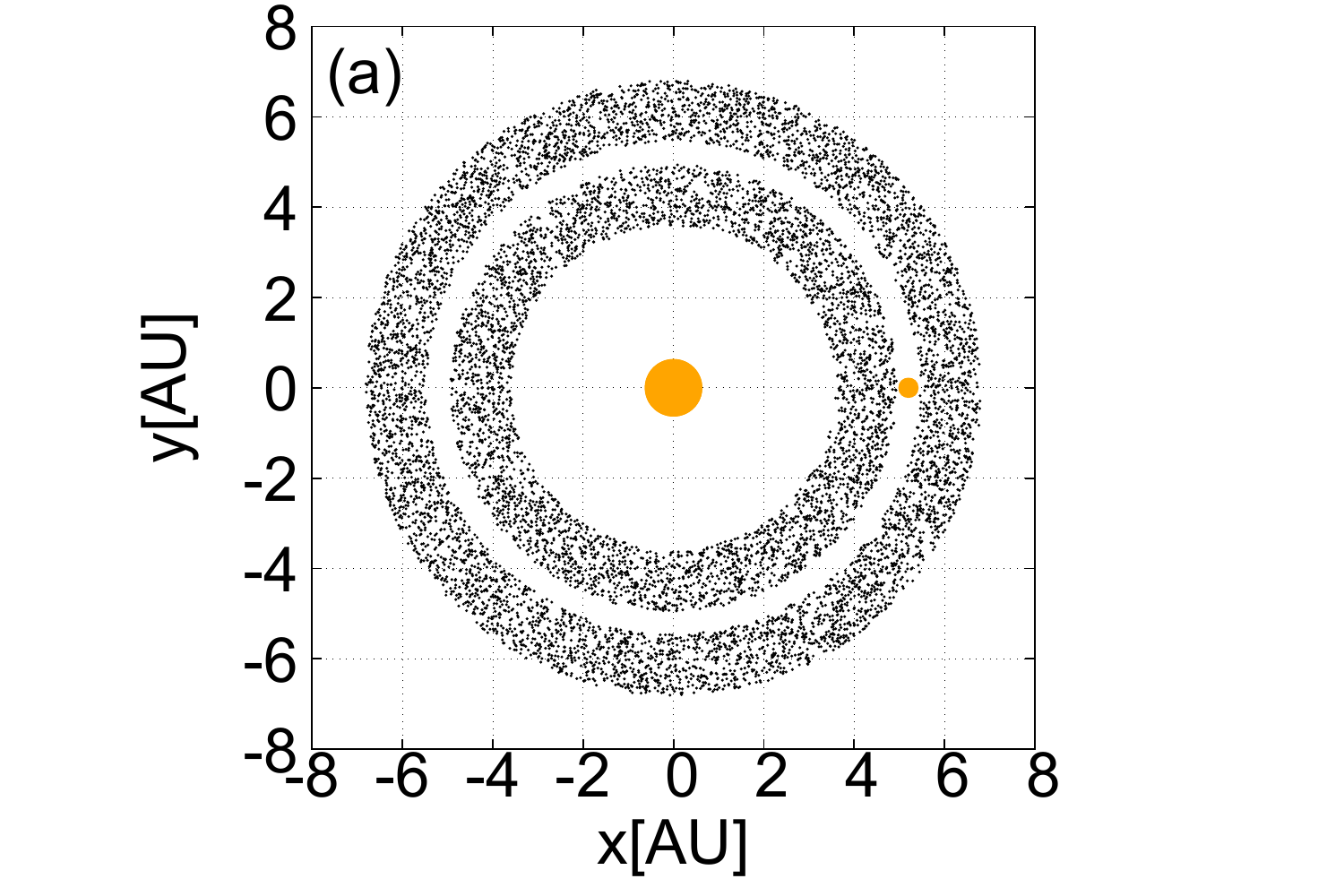}
    \includegraphics[width=80mm]{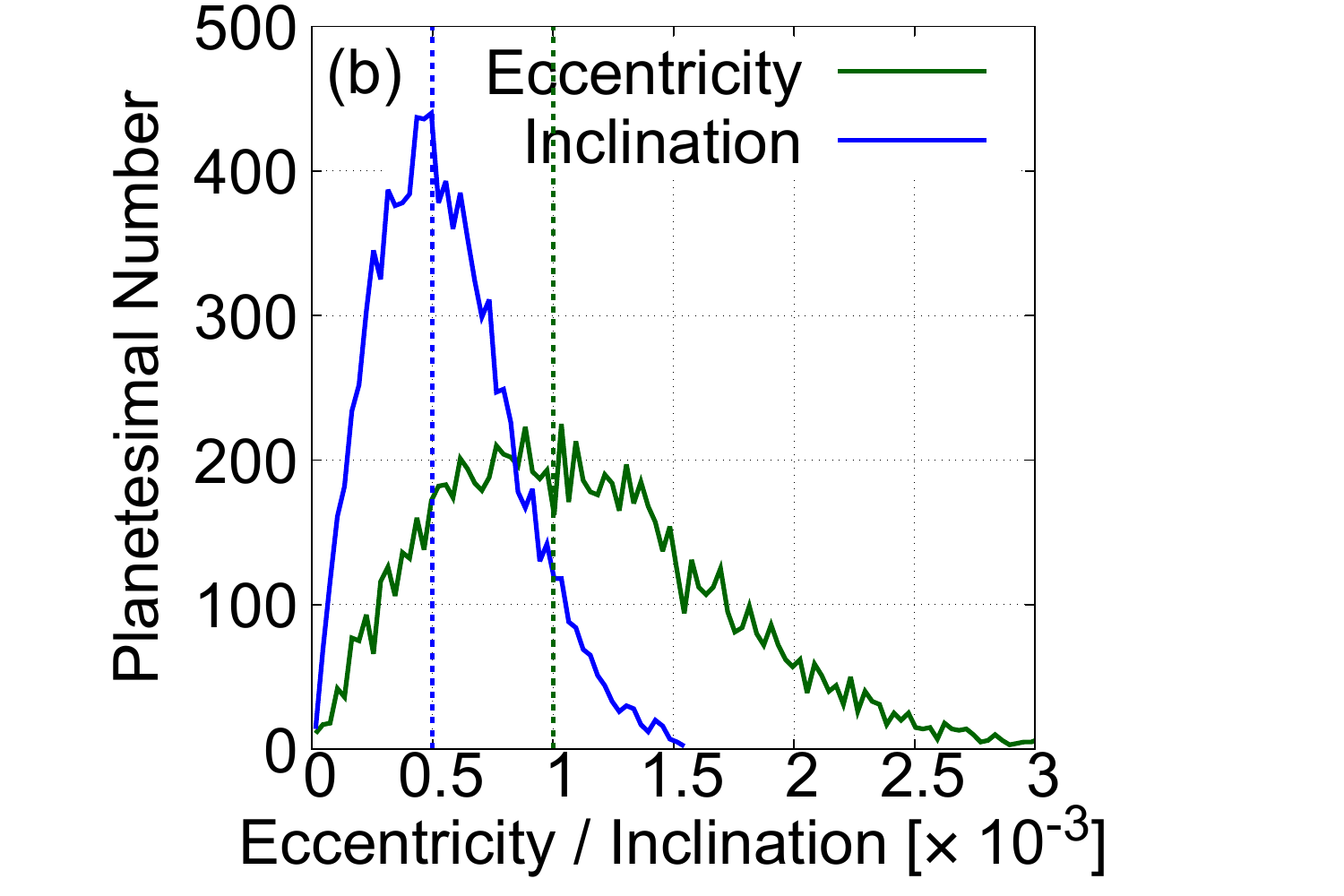}
    \caption{
    Initial distribution of planetesimals: 
    Panel ($a$) shows the spatial distribution in the $x$-$y$ plane. 
    The big orange filled circle at the origin indicates the central star, while the small one is the protoplanet. 
    Planetesimals are uniformly distributed, except for the feeding zone of the protoplanet of its initial mass. 
    Panel ($b$) shows the initial number of planetesimals as functions of eccentricity (green) and inclination (blue), which are generated from the probability distribution function, \refeqs{eq:app_dist_ecc_inc}.
    Here we assume ${\statave{{\ecc}^{2}}}^{1/2} = 0.001$ and ${\statave{{\inc}^{2}}}^{1/2}=0.0005$.    
    }
    \label{fig:Initial}
    \end{center}
\end{figure}

As described above,
the protoplanet is assumed to grow only via gas accretion from  its initial mass $\Mpint$ to final mass $\Mpfnl$.
We set $\Mpint = 8.0 \times 10^{-5} \Msun$ $(\lesssim 30~\Mearth)$ and $\Mpfnl = 1.0 \times 10^{-3} \Msun$ $(\gtrsim 1 \Mjupiter)$.
At the beginning of simulations, we assume that planetesimals that existed in the feeding zone of the protoplanet of mass $\Mpint$
\revised{
, inside which planetesimals have a possibility to be captured by the protoplanet (see Section~3.1 for the exact definition),
}
have already accreted in the protoplanet. 
Thus, we distribute planetesimals initially between the semi-major axes $\axi$ = $3.6~\AU$ ($\equiv \axiin$) and $6.8~\AU$ ($\equiv \axiout$), which covers the feeding zone of the protoplanet of mass $\Mpfnl$, except for the zone between $4.7~\AU$ ($\equiv \axiinzero$) to $5.7~\AU$ ($\equiv \axioutzero$) (equivalent to the initial feeding zone). 
The initial distribution is visualized in Fig.~\ref{fig:Initial}$a$.

We set the initial values of the Keplerian orbital elements of planetesimals, semi-major axis $\axi$, eccentricity $\ecc$, inclination $\inc$, longitude of ascending node $\lan$, longitude of pericenter $\lpc$ and true anomaly $\tra$ in the following way. 
The initial total mass of planetesimals is equal to that of solids in the zones $\axiin \leq \axi \leq \axiinzero$ and $\axioutzero \leq \axi \leq \axiout$ in the minimum-mass solar nebula \citep[MMSN;][]{Hayashi1981}, namely $6.7 \Mearth$ ($\equiv \Mdust$). 
Note that a particle that we call a planetesimal in this study is, in fact, not a real planetesimal, but an imaginary super-particle that represents a group of many identical planetesimals. 
When simulating its orbital motion, we regard the particle as a real planetesimal of radius $\Rplts$ and mass $\Mplts$, while we regard it as a super-particle when estimating the total mass of captured planetesimals.
The super-particles have the same mass $\Mdust/\Nplts$ ($\Nplts$ being the total number of the super-particles = 10000) and the same radius $\Rplts$.
The semi-major axis of the particle labeled as $j$ ($j$ = 1, $\cdots$, $\Nplts$), $\axi_j$, is given in such a way that
\begin{equation}
	\frac{j}{\Nplts} = \frac{1}{\Mdust} \int_{\axiin}^{\axi_j} \Sigmadust \cdot 2 \pi \axi \, {\rm d} \axi,
\end{equation}
so that the mass distribution is consistent with the surface density of solids $\Sigmadust$, which is assumed to be spatially uniform in this study. 

As for eccentricity and inclination, assuming that planetesimals are scattered by their mutual gravitational interaction, we adopt the Rayleigh distribution given by \citep{Lissauer1993}
\begin{equation}\label{eq:app_dist_ecc_inc}
	\fei{\ecc}{\inc} = 4 \frac{\Sigmadust}{\Mplts} \frac{\ecc \inc}{\statave{{\ecc}^{2}} \statave{{\inc}^{2}} } \exp{ \left[ - \frac{{\ecc}^2}{\statave{{\ecc}^{2}}} - \frac{{\inc}^2}{ \statave{{\inc}^{2}} } \right] },
\end{equation}
as the initial $\ecc$ and $\inc$ of planetesimals. 
We set ${\statave{{\ecc}^{2}}}^{1/2}=0.001$ and ${\statave{{\inc}^{2}}}^{1/2}=0.0005$; however, choice of $\statave{\ecc^2}$ and $\statave{\inc^2}$ scarcely affects our conclusions because viscous stirring by the protoplanet changes the distribution immediately.
The orbital angles $\lan$, $\lpc$ and $\tra$ are distributed uniformly.
Finally, we transform these orbital elements into the Cartesian coordinate and then perform orbital integration. 

\subsection{Summary of case studies}
The main focus of this study is on the effects of gap formation on capture of planetesimals by a growing protoplanet. 
We investigate such effects in a step-by-step fashion. 
The free parameters include $\Sigmaf$, $\Mdotplanet$, and $\Rplts$.
Firstly, we investigate the case where $\Sigmaf$ = 1 (i.e., no gap) and $\Mdotplanet$ is constant with time. 
This model is regarded as the standard one and referred to as \Mupd ($\refsec{sec:Model1}$).
Secondly, we consider the effects of gap formation with constant $\Mdotplanet$ (\Mcdt; $\refsec{sec:Model2}$).
Thirdly, we explore the case where the protoplanet embedded in a gapped disc grows with a mass-dependent rate (\ModelC; $\refsec{sec:Model3}$).
For $\Rplts$ we consider four values, $1 \times 10^4,$ $10^5$, $10^6$, and $10^7 \cm$ in each model.
The cases and parameter values we consider are summarized in \reftab{tb:settings}.

\begin{table*}
	\centering
        {\scriptsize
	\begin{tabular}{ llllllll } 
		\hline
		Model 		& $\Sigmaf$ 			    & $\Mdotplanet$ 	    & $\etadisk$           & $(\alphaPPD, \taupep)$         & Free Parameters                   & used in &\\
		\hline
		$\Mupd$ 	& 1 				        & constant			    & \refeqs{eq:eta_disk} & -                                      & $\Rplts,\Mdotplanet$              & Sec. \ref{sec:Model1} &\\
		$\Mgwoe$ 	& \refeqs{eq:Method_Sigma} 	& constant			    & 0                    & -                                      & $\Rplts,\Mdotplanet,\alphaPPD$    & Sec. \ref{sec:Model2} &\\
		$\Mgwe$ 	& \refeqs{eq:Method_Sigma} 	& constant			    & \refeqs{eq:eta_disk} & -                                      & $\Rplts,\Mdotplanet,\alphaPPD$    & Sec. \ref{sec:Model2} &\\
		$\ModelCa$ 	& \refeqs{eq:Method_Sigma}  & \refeqs{eq:ACC_Rate}  & \refeqs{eq:eta_disk} & $(\infty,  \infty)$                & $\Rplts$                          & Sec. \ref{sec:Model3} &\\
		$\ModelCb$ 	& \refeqs{eq:Method_Sigma}  & \refeqs{eq:ACC_Rate}  & \refeqs{eq:eta_disk} & $(10^{-2}, 10^{3})$  & $\Rplts$                          & Sec. \ref{sec:Model3} &\\
		$\ModelCc$ 	& \refeqs{eq:Method_Sigma}  & \refeqs{eq:ACC_Rate}  & \refeqs{eq:eta_disk} & $(10^{-3}, 10^{4})$  & $\Rplts$                          & Sec. \ref{sec:Model3} &\\
		$\ModelCd$ 	& \refeqs{eq:Method_Sigma}  & \refeqs{eq:ACC_Rate}  & \refeqs{eq:eta_disk} & $(10^{-4}, 10^{5})$  & $\Rplts$                          & Sec. \ref{sec:Model3} &\\
		\hline
	\end{tabular}
	}
	\caption{
	Model settings.
        }
    \label{tb:settings}
\end{table*}

\section{Capture of Planetesimals in a Uniform Circumstellar Disc}
\label{sec:Model1}
\subsection{\revised{Basic Mechanism for Orbital Evolution of Planetesimals around a Growing Protoplanet}}\label{sec:Result_1_orb}

\begin{figure}
    \begin{center}
    \includegraphics[width=65mm]{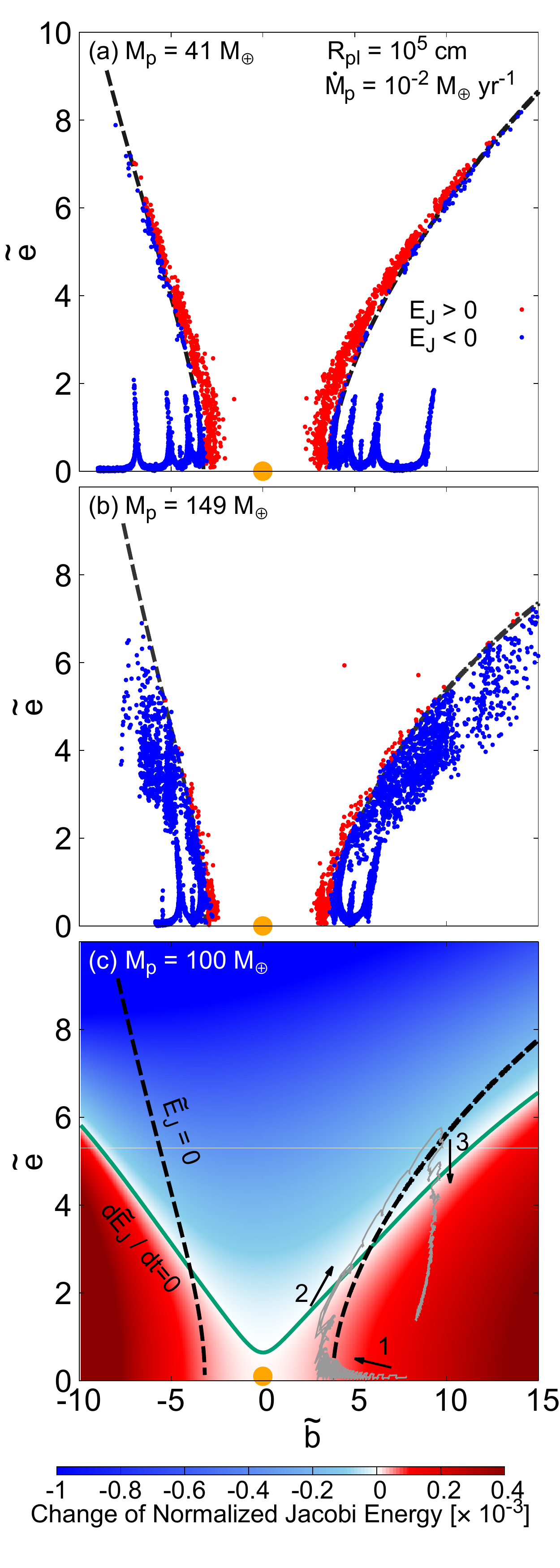}
    \caption{
    Snapshots of the orbital evolution of planetesimals around a growing protoplanet of mass 41~$\Mearth$ (panel (a)) and 149~$\Mearth$ (panel (b))
    on the plane of $\bnrm$ and $\eccnrm$ (see \refeqs{eq:def_eccnrm} for their definitions) for the unperturbed disc model(Model A).
    The orange filled circle at the origin represents the protoplanet.
    The red and blue symbols represent planetesimals with the Jacobi energy $\Ejacobi > 0$ and $\Ejacobi < 0$, respectively.
    The dashed lines indicate the outer boundary of the feeding zone \revised{(i.e., $\Ejacobi = 0$; written by \refeqs{eq:E_Jacobi_orb2} with $\inc=0$)}. 
    Panel~(c) is a contour plot of the temporal change in Jacobi energy $\Ejdotnrm$ on the $\bnrm$-$\eccnrm$ plane for the protoplanet mass $\Mplanet = 100 \Mearth$. In the red and blue domains, $\Ejdotnrm$ is positive and negative, respectively.
    The green solid line is for $\Ejdotnrm$ = 0 while the dotted lines are for $\tilde{\Ejacobi}$ = 0. 
    The gray solid line shows the evolution path of a planetesimal around a growing protoplanet.
    In all those panels, the planetesimal size $\Rplts$ is $1~\times~10^5~\cm$ and the protoplanet growth rate $\Mdotplanet$ is $1~\times~10^{-2}~\Mearth~{\yr}^{-1}$.
    }
    \label{fig:ResultsORB_be}
    \end{center}
\end{figure}


Before investigating the effects of a gap in the circumstellar disc on the capture of planetesimals by a growing protoplanet, we review the numerical results and analytical interpretation given by \citet{Shiraishi+2008}, which is helpful for interpreting our numerical results. 
According to their analysis, the surface density of planetesimals in the so-called feeding zone is the most important factor that determines the capture rate of planetesimals during late stages of gas giant formation.
Here, we show the relation between the capture rate and surface density of planetesimals, based on \citet{Shiraishi+2008}.
Then, we give an interpretation to the change in the surface density of planetesimals in terms of the Jacobi energy.

The total mass of planetesimals captured per unit time by the protoplanet (called simply the capture rate, hereafter) is given by
\begin{align}\label{eqs:accretion_rate}
    \Mdotcap = \Mplts \nplts \collsec \vrel,
\end{align}
where $\nplts$ is the number density of planetesimals, $\collsec$ is the collision cross section of the protoplanet and $\vrel$ is the planetesimal's velocity relative to the protoplanet, which depends on its eccentricity and inclination. 
$\collsec$ depends on whether the capture radius $\Rcap$ is larger or smaller than the scale height of the planetesimal disc $\hscaleplts$.
The former is defined as the threshold value of the impact parameter below which planetesimals can collide the protoplanet and is given with $\vesc$ being the escape velocity from the protoplanet by 
\begin{align}\label{eq:capture_radius}
    \Rcap = \Rplanet \left\{ 1 + \left( \frac{\vesc}{\vrel} \right)^2 \right\}^{1/2},
\end{align}
provided the random velocity is high enough that the stellar gravity is negligible.
Although not verified a priori, this condition is approximately achieved in the numerical simulations of \citet{Shiraishi+2008} and ours.
For a massive protoplanet of interest in this study ($> 1 \Mearth$), $\vesc > \vrel$ and thus $\Rcap \sim \Rplanet \vesc / \vrel$ from Eq.~(\ref{eq:capture_radius}).
Also, $\hscaleplts \sim \axiplanet \inc$ and $\vrel \sim \ecc \vKep$. 
Thus, the condition $\Rcap > \hscaleplts$ is written in terms of protoplanetary mass as
\begin{equation}
    \frac{\Mplanet}{\Mstar} > \frac{1}{2} \frac{\axiplanet}{\Rplanet} \ecc^2 \inc^2,
\end{equation}
which is estimated to be
\begin{align}\label{eqs:condition_for_2d_regime}
    \frac{\Mplanet}{\Mstar} 
                            &> 6.26 \times 10^{-8}  \left( \frac{\rhoave}{1.25 \times 10^{-1} \g {\cm}^{-3}} \right)^{1/4} 
                                                    \left( \frac{\Mstar}{\Msun} \right)^{-1/3} \nonumber \\
                            & \quad             \left( \frac{\axiplanet}{5.20 \AU} \right)^{3/4}
                                                \left( \frac{\ecc}{1.00 \times 10^{-3}} \right)^{3/2}
                                                \left( \frac{\inc}{1.00 \times 10^{-3}} \right)^{3/2} \revised{.}
\end{align}
This condition always holds true in this study, so that 
\revised{
the collision cross section is regarded approximately as a rectangle with height of $2 \hscale$ and width of $2 \Rcap$, namely
}
$\collsec \simeq 4 \Rcap \hscaleplts$.
Thus, Eq.~(\ref{eqs:accretion_rate}) becomes
\begin{align}\label{eq:accretion_rate2}
	\Mdotcap 	&\sim  \frac{\Sigmadust}{2 \hscaleplts} \cdot 4 \Rcap \hscaleplts \cdot \vrel \nonumber \\
	            &\sim \sqrt{ \frac{32 \pi \Gconst \rhoave}{3} } {\Rplanet}^2 \Sigmadust.
\end{align}
This indicates that $\Mdotcap$ depends only on $\Sigmadust$ for given $\rhoave$ and $\Rplanet$.
Although $\rhoave$ and $\Rplanet$ also vary during protoplanetary growth, here we focus on the physical process that determines $\Sigmadust$.

According to \citet{Shiraishi+2008}, using the surface density of planetesimals averaged over the feeding zone as $\Sigmadust$ in Eq.~(\ref{eq:accretion_rate2}) (hereafter, called the effective surface density and denoted by $\Sigmaeff$), one obtains an estimate for $\Mdotcap$ that is consistent with their numerical results. 
The feeding zone is defined with the Jacobi energy $\Ejacobi$ as \citep[e.g.][]{Hayashi+1977},
\begin{align}
   \Ejacobi &\equiv \frac{1}{2} {\vpltsp}^2 + \Ujacobi > 0, \label{eq:Jacobi_Energy}
\end{align}
where $\Ujacobi$ is the Jacobi potential defined as
\begin{align}
   \Ujacobi &= - \frac{1}{2} {\Omegap}^2 \left( {{x}^{\prime}}^2 + {{y}^{\prime}}^2 \right) - \Gconst \frac{\Mstar}{\rstpl} - \Gconst \frac{\Mplanet}{\rplpl} + U_0; \label{eq:Jacobi_Potential}
\end{align}
$\vpltsp$ and ($x^\prime, y^\prime$) are the velocity and position of the planetesimal, respectively, in the coordinate system co-rotating with the protoplanet, 
and $U_0$ is a constant determined according to the boundary condition. 
Here, we set $U_0$ such that $\Ujacobi$ vanishes at the Lagrange L$_2$ point and thus
\begin{align}\label{eq:U_Jacobi_0}
	U_0 = \frac{\Gconst \left( \Mstar + \Mplanet \right) }{ \axiplanet } \left[ \frac{3}{2} + \frac{9}{2} {\hill}^2 + O \left( {\hill}^3 \right) \right].
\end{align}
By definition, $\Sigmaeff$ is controlled by the change in the Jacobi energy.
Below, we give an interpretation to the orbital evolution of planetesimals around the growing protoplanet in terms of the Jacobi energy.

First, in \reffig{fig:ResultsORB_be}, we show snapshots of the orbital evolution of planetesimals around the growing protoplanet at $\Mplanet$ = (a) 41~$\Mearth$ and (b) 149~$\Mearth$ for \Mupd.
Here, the horizontal and vertical axes are the normalized semi-major axis $\bnrm$ and eccentricity $\eccnrm$, which are defined respectively as
\begin{equation}\label{eq:def_eccnrm}
	\bnrm 	\equiv \frac{\axi - \axiplanet}{ \axiplanet \hill}, 
	\hspace{3ex}
	\eccnrm  \equiv \frac{\ecc}{\hill}, 
\end{equation}
where $\hill$ is the reduced Hill radius defined as
\begin{equation}\label{eq:reduced_hill}
	\hill \equiv \left( \frac{\Mplanet}{3 \Mstar} \right)^{1/3}.
\end{equation}
The growth rate $\Mdotplanet$ is $1~\times~10^{\revised{-2}}~\Mearth~{\yr}^{-1}$ and the size of planetesimals $\Rplts$ is $1 \times 10^5~\cm$.
Planetesimals are color-coded according to their values of $\Ejacobi$ and the red and blue symbols represent $\Ejacobi > 0$ and $< 0$, respectively.
The black dashed lines indicate the outer boundary of the feeding zone: 
Using the orbital elements and setting $\Mstar$, $\axiplanet$ and $\TKep (\axiplanet) / 2 \pi$ as the units of mass, length and time, respectively, \refeqs{eq:Jacobi_Energy} can be transformed as \citep[e.g.][]{Hayashi+1977} 
\begin{align}\label{eq:E_Jacobi_orb2}
	\Ejacobi = - \frac{1}{2 \axi} - \sqrt{ \axi \left( 1 - {\ecc}^2 \right) } \cos \inc + \frac{3}{2} + \frac{9}{2} {\hill}^2 + O \left( {\hill}^3 \right)
\end{align}
or
\begin{align}\label{eq:E_Jacobi_orb3}
	\Ejacobinrm \equiv \frac{\Ejacobi}{ {\hill}^2 } = \frac{1}{2} \left( {\eccnrm}^2 + {\incnrm}^2 \right) - \frac{3}{8} {\bnrm}^2 + \frac{9}{2} + O \left(\baxi,\ecc,\inc,\hill \right),
\end{align}
where $\incnrm$ is the normalized inclination ($\equiv \inc / \hill$).
For \revised{$b, \ecc, \inc, h \ll 1$}, the shape of the feeding zone is fixed on the $\bnrm$--$\eccnrm$ plane, as shown in \reffig{fig:ResultsORB_be}.

The dynamic behaviour of planetesimals shown in \reffig{fig:ResultsORB_be} is outlined as follows.
Outside the feeding zone, except for mean motion resonances to the protoplanet, the normalised eccentricities of planetesimals are much smaller than unity.
This is because gravitational scattering by the protoplanet is so week far from the protoplanet that the gas drag keeps the planetesimals' eccentricities small.
At the mean motion resonances, the eccentricities are relatively high because of amplified gravitation scattering.
Inside the feeding zone, the planetesimals experience close encounters with the protoplanet and are scattered into highly eccentric orbits.
Since the gravitational scattering itself never changes the Jacobi energy, it merely distributes planetesimals along the constant Jacobi energy line and does not result in ejection from nor injection into the feeding zone.

The planetesimals go into the feeding zone through the protoplanet growth and get farther \revised{away} from the feeding zone through \revised{the damping and/or orbital decay by disc gas drag.}
This is explained in terms of change in the Jacobi energy, which is expressed as
\begin{align}\label{eq:der_E_Jacobi_orb}
	\deriv{\Ejacobinrm}{t} = \frac{1}{{\hill}^2} \deriv{\Ejacobi}{t} - \frac{2}{3} \frac{\Ejacobi}{{\hill}^2} \frac{1}{\tauacc},
\end{align}
where $\tauacc$ ($\equiv \Mplanet / \Mdotplanet$) is the characteristic timescale of protoplanet growth and
\begin{align}\label{eq:der_E_Jacobi_orb2}
	\deriv{\Ejacobi}{t} 	&=&  {\axi}^{1/2}  \left[ \frac{1}{2 \axi} \deriv{\axi}{t} \left( {\axi}^{-3/2} - \sqrt{ 1 - {\ecc}^2 } \cos \inc \right) \right. \nonumber \\
				 	&&	\left. 	+ \frac{1}{\ecc} 	\deriv{\ecc}{t} 	\frac{ {\ecc}^2 }{ \sqrt{ 1 - {\ecc}^2 } } \cos \inc \right. \nonumber. \\
					&&	\left.	+ \frac{1}{\inc} 	\deriv{\inc}{t} 	\sqrt{ 1 - {\ecc}^2 } \inc \sin \inc \right] \nonumber  \\
					&& + 3 {\tauacc}^{-1} {\hill}^2,
\end{align}
from \refeqs{eq:E_Jacobi_orb2}.
The characteristic timescales of change in $\axi$, $\ecc$ and $\inc$ via gas drag are given for constant $\cd$ as \revised{\citep{Adachi+1976,Kary+1993}}
\begin{align}
	\taudampaxi \equiv a \left|\deriv{\axi}{t}\right|^{-1} &= - \frac{\taudampzero}{2} \left\{ 
		\left( 0.97 \ecc + 0.64 \inc + |\etadisk| \right) \etadisk 
		\right\}^{-1}, \label{eq:def_taudamp_axi}\\
	\taudampecc \equiv \ecc \left|\deriv{\ecc}{t}\right|^{-1} &= - \taudampzero  \left\{ 
		0.77 \ecc + 0.64 \inc + \revised{1.5}|\etadisk| 
		\right\}^{-1}, \label{eq:def_taudamp_ecc}\\
	\taudampinc \equiv \inc \left|\deriv{\inc}{t}\right|^{-1} &= - \taudampzero  \left\{ 
		0.77 \ecc + 0.85 \inc + |\etadisk| 
		\right\}^{-1}, \label{eq:def_taudamp_inc}		
\end{align}
where
\begin{equation}\label{eq:def_taudamp}
	\taudampzero = \frac{2 \Mplts}{\cd \pi \Rplts^2 \rhogas \vKep}.
\end{equation}
Damping of eccentricity and inclination due to gas drag results in a decrease in the Jacobi energy. 
Change in the semi-major axis results in increase or decrease in the Jacobi energy depending on the disc gas flow ($\etadisk$) and orbital elements.
When $\cd$ is not constant, we have to integrate \refeqs{eq:der_E_Jacobi_orb} for one orbital period directly to know the change of Jacobi energy; however, the qualitative characteristics concerning the changing rate of the Jacobi energy are almost the same.

Panel (c) of \reffig{fig:ResultsORB_be} shows the averaged values of $\Ejdotnrm$ on the $\bnrm$--$\eccnrm$ plane for $\Mplanet = 100 \Mearth$\revised{, which are obtained by integrating \refeqs{eq:der_E_Jacobi_orb} for one orbital period with variable $\cd$.
The detailed method of the integration is shown in the \refapp{App_Ejacobi}.
}
Planetesimals in the red area have their Jacobi energies enhanced (i.e., $\Ejdotnrm > 0$) and move toward the feeding zone, while their Jacobi energies decreased (i.e., $\Ejdotnrm < 0$) in the blue area. 
At the boundary (i.e., $\Ejdotnrm=0$) indicated by the green solid line, the growth and damping timescales are similar to each other.
The gray solid line shows the evolution path of a planetesimal.
From this panel, one can predict how those planetesimals move around the growing protoplanet: 
Far from the protoplanet, since their equilibrium eccentricities are sufficiently small, the planetesimals stay in the red area and their Jacobi energies tend to increase.
This means that those planetesimals are pushed toward the feeding zone(array 1).
Once they enter the feeding zone ($\Ejacobi > 0$), their eccentricities become further higher due to strong gravitational scattering by the protoplanet(array 2). 
Many of the planetesimals move into the blue area and their Jacobi energies decrease. 
This means that the planetesimals are pushed out of the feeding zone(array 3).

The above analysis indicates that the capture of planetesimals depends on the effective surface density of planetesimals $\Sigmaeff$ and also on the rate of change in the Jacobi energy, which depends on the rates of protoplanetary growth and damping of planetesimals' velocity dispersion. 
The knowledge of the relation among the capture rate of planetesimals and these timescales is useful to understand the numerical results shown in succeeding sections.

\subsection{Numerical Results}
\begin{figure}
    \begin{center}
    \includegraphics[width=80mm]{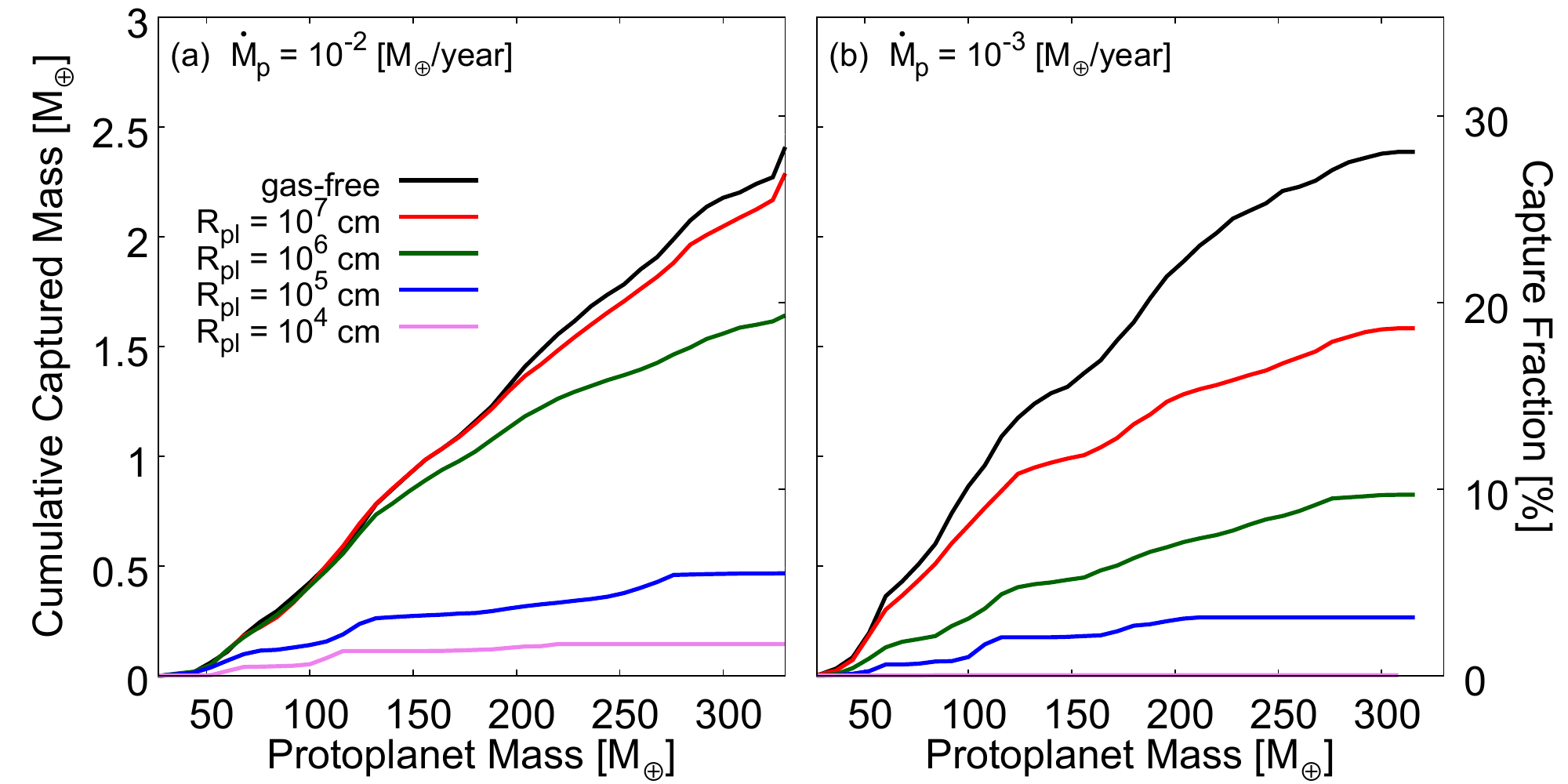}
    \caption{
    Temporal change in the cumulative mass of planetesimals captured by the protoplanet in the unperturbed disc model ($\Mupd$).
    In panel (a) and (b), the protoplanetary growth rate $\Mdotplanet$ is $1 \times 10^{-2}$~$ \Mearth / \yr$ and $1 \times 10^{-3}$~$ \Mearth / \yr$, respectively.
    The left-$y$ axis is the absolute captured mass, while the right-$y$ axis is the mass relative to the initial total mass of planetesimals (called the capture fraction).
    The $x$ axis is the protoplanet mass, which is equivalent to time.
    The lines are color-coded according to the size of planetesimals $\Rplts$, as indicated in the panel, except for the black line which represents the case without gas drag.
    }
    \label{fig:ModelA_Mcap}
    \end{center}
\end{figure}

\revised{Here, we investigate the capture of planetesimals in a uniform circumstellar disc (\Mupd; see Table~\ref{tb:settings}).}
Figure~\ref{fig:ModelA_Mcap} shows the temporal changes in cumulative mass of planetesimals captured by the protoplanet (simply the cumulative captured mass or $\Mcap$, hereafter) with protoplanetary growth for $\Mdotplanet$ = $1 \times 10^{-2}$~$\Mearth~{\yr}^{-1}$ in panel (a) and for $\Mdotplanet$ = $1 \times 10^{-3}$~$\Mearth~{\yr}^{-1}$ in panel (b). 
The black line represents the \revised{the gas-free} case and the coloured lines represent the cases of different planetesimal size.

First, we investigate the dependence of the cumulative captured mass $\Mcap$ on gas drag or, equivalently, planetesimal size in the higher-$\Mdotplanet$ case (Fig.~\ref{fig:ModelA_Mcap}a). 
Planetesimals are captured most efficiently in \revised{the gas-free case.}
In the case of $\Rplts$ = $1~\times~10^7~\cm$ for which gas drag is less effective, the cumulative captured mass increases quite similarly until $\Mplanet = 200~\Mearth$, but more slowly for $\Mplanet > 200~\Mearth$ than in \revised{the gas-free case.}
As planetesimal size decreases, the coloured line starts to deviate from the black line at a smaller mass, that is, $\Mplanet = 130~\Mearth$ for $\Rplts$ = $1~\times~10^6~\cm$, while $\Mplanet = 60~\Mearth$ for $\Rplts$ = $1~\times~10^5~\cm$.
In the case of $\Rplts$ = $1~\times~10^4~\cm$, the cumulative captured mass always increases more slowly than that in \revised{the gas-free case.}
Thus, the total captured mass ($\Mcaptot$: $\Mcap$ at $\Mplanet = \Mpfnl$) decreases with planetesimal size.

In the lower-$\Mdotplanet$ case (Fig.~\ref{fig:ModelA_Mcap}b), the cumulative captured mass is smaller than in the higher-$\Mdotplanet$ case, except for \revised{the gas-free case} where the total captured mass is independent of $\Mdotplanet$.
In the case of $\Rplts$ = $1~\times~10^4~\cm$, no planetesimals are captured by the protoplanet.

Such dependence can be interpreted, in terms of the Jacobi energy, as follows.
In \revised{the gas-free case}, since the damping timescale is infinity, one obtains
\begin{align}\label{ModelA_dEj_taudamp_inf}
    \deriv{\Ejacobinrm}{\Mplanet}   &= \frac{1}{\Mplanet} \left(3 - \frac{2}{3} \Ejacobinrm \right),
\end{align}
from Eqs.~(\ref{eq:der_E_Jacobi_orb}) and (\ref{eq:der_E_Jacobi_orb2}).
The above equation indicates that the Jacobi energy keeps increasing up to $\Ejacobinrm = 9/2$ and final Jacobi energy which a planetesimal has at $\Mplanet=\Mpfnl$ is determined only by the initial and final conditions of the integration.
The growth rate $\Mdotplanet$ never affects the value, which means
how many planetesimals enter the feeding zone depends on $\Mpfnl$, but not on  $\Mdotplanet$.
That is the reason why the total captured mass takes the same value regardless of $\Mdotplanet$.

When gas drag is effective, some of the planetesimals inside the feeding zone are pushed out before being captured by the protoplanet.
When the protoplanet is small in mass, its growth timescale is so short relative to the damping timescale that the planetesimals keep staying in the feeding zone and $\Sigmaeff$ is almost the same as that in \revised{the gas-free case.}
Then, the protoplanet grows and thereby the growth timescale becomes long enough that the gas drag is effective, reducing $\Sigmaeff$ and thereby making the cumulative captured mass smaller relative to \revised{the gas-drag case.}
How much $\Sigmaeff$ is reduced depends on those two timescales, because ${\rm d} {\tilde E}_{\rm J} / {\rm d}t$ also depends on them.
In summary, the cumulative captured mass becomes smaller for smaller planetesimals or slower protoplanet growth.

\subsection{Comparison with  Analytical Formulae from \citet{Shiraishi+2008}}\label{sec:Discussion_SI}
As described in Introduction, the previous study, \citet{Shiraishi+2008}, investigated the capture of planetesimals in the unperturbed disc and derived semi-analytical formulae for the capture rate of planetesimals that explain their numerical results.
\revised{Here, we compare our results with the analytical formulae from \citet{Shiraishi+2008}.}

They used a simplified gas drag formula,
\begin{align}\label{eqs:ModelA_SI_Drag}
	{\bf \fgas} = - \frac{\bf \vplgs}{\taudamp},
\end{align}
where the damping timescale $\taudamp$ was assumed to be unchanged through their simulations. 
Also they assumed that the disc gas rotates around the central star with the Kepler velocity, namely $\etadisk=0$.
They found that the capture process can be divided into two different regimes.
The first regime is such that no clear gap is formed in the swam of planetesimals that surrounds the protoplanet (called the {\it planetesimal disc}, hereafter). 
This regime is achieved when $\tauacc < \taudamp$, which means the protoplanet's growth dictates the capture of planetesimals. 
Thus, $\Mdotcap$ is approximately determined only by $\tauacc$.
The second regime in which $\tauacc > \taudamp$ is such that a clear gap is formed in the planetesimal disc.
In this regime, $\Mdotcap$ depends not only on $\tauacc$ but also on $\taudamp$.
The second regime is, therefore, affected by assumptions regarding the structure of the surrounding gas disc.

\begin{figure}
    \begin{center}
    \includegraphics[width=80mm]{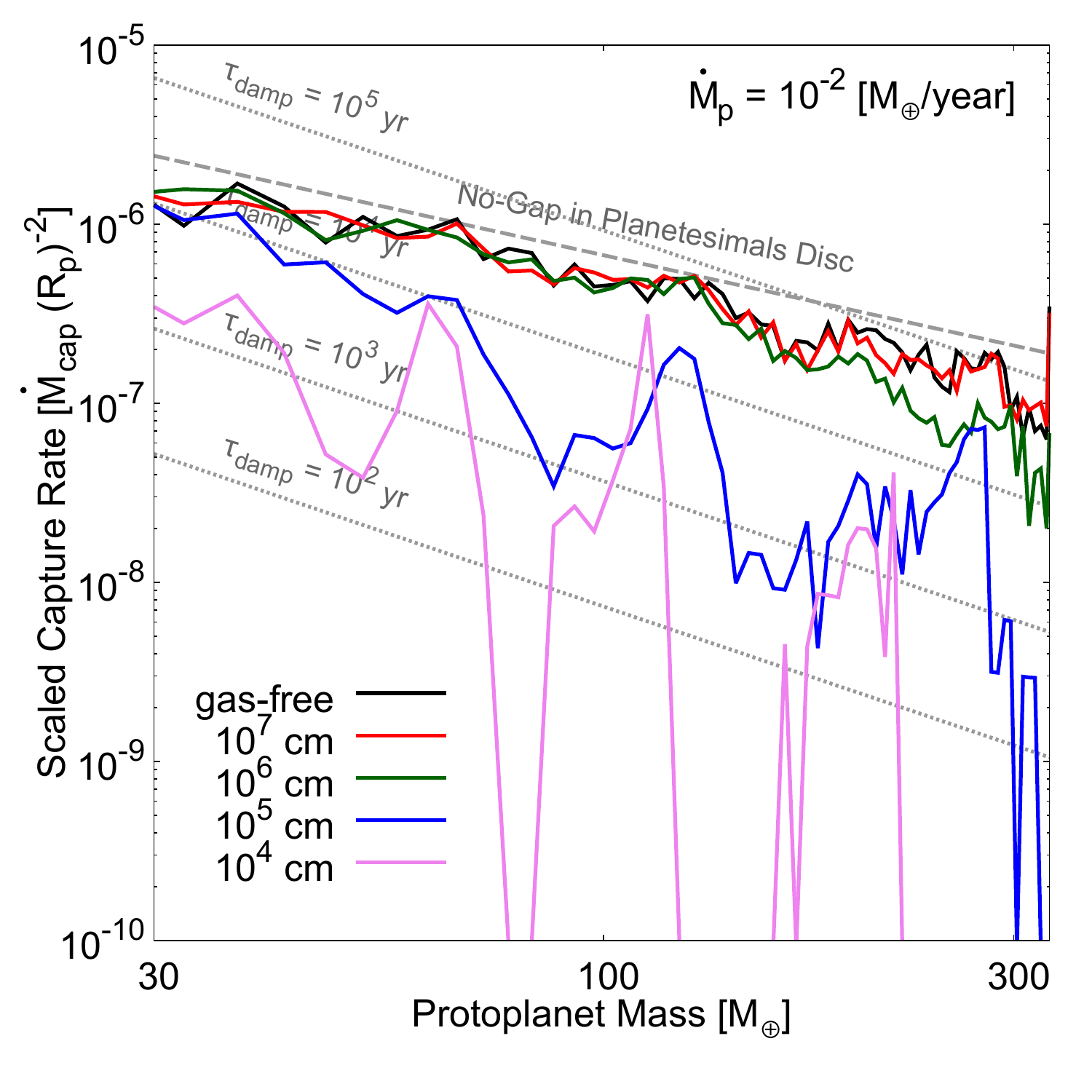}
    \caption{
    Change in the capture rate of planetesimals with protoplanet growth.
	The capture rate, $\Mdotcap$, scaled by the square of the protoplanet's radius, $\Rplanet$, is plotted as a function of the protoplanet's mass. 
	Our results for the unperturbed disc are shown by solid lines colour-coded according to the assumed size of planetesimals.
	The gray dashed and dotted lines represent the semi-analytical formulae for the planetesimal disc without and with a gap, respectively, derived by \citet{Shiraishi+2008}.
    The adopted values of $\taudamp$ are indicated.
    }
    \label{fig:Discussion_SI_1}
    \end{center}
\end{figure}

Figure~\ref{fig:Discussion_SI_1} shows the change in $\Mdotcap$ with the protoplanetary growth.
The semi-analytical formula for the first regime is shown by the gray dashed line, and those for the second regime with different values of $\taudamp$ are shown by the gray dotted lines.
Firstly, in the cases of no drag and the largest planetesimals ($\Rplanet = 1 \times 10^7$~cm), the black and red lines are almost parallel to the gray dashed one, since $\tauacc < \taudamp$ (i.e., the first regime).
Secondly, the green and blue lines ($\Rplanet = 1 \times 10^6$ and $1 \times 10^5$~cm, respectively) are initially similar to the gray dashed one.
As the protoplanet grows, its growth timescale ($\propto \Mplanet$) becomes longer and the regime switches to the other one.
Indeed, the green and blue lines depart from the dashed line at $\Mplanet \sim 130 \Mearth$ and $60 \Mearth$, respectively, and shift toward the lines for the second regime.
Finally, as for the smallest planetesimals ($\Rplanet = 1 \times 10^4$~cm; magenta line), the capture process is in the second regime from the beginning on.
It turns out that the numerical results are considerably different from the semi-analytical formulae in the second regime: the calculated $\Mdotcap$ decreases with $\Mplanet$ more rapidly than that given by the semi-analytical formulae and show a jagged profile.
The difference between the numerical and semi-analytical $\Mdotcap$ is more striking for smaller planetesimals.

Obviously, such inconsistency between the numerical and semi-analytical results comes from the different treatments for gas drag. 
Our simulations demonstrate that both \revised{the variable damping timescale of planetesimals' eccentricities} and the sub-Kepler rotation of the disc gas (i.e., radial drift of planetesimals), which \citet{Shiraishi+2008} ignored, have great impacts on the capture rate of planetesimals.
As shown in the next sections, if a gap is formed in the \textit{gas} disc, both the timescales of eccentricity damping and radial drift are significantly changed and, thus, the condition of $\tauacc > \taudamp$ is readily achieved.
It will turn out that the capture rate of planetesimals changes more complexly with protoplanet growth and is significantly different from prediction by the semi-analytical formulae derived in \citet{Shiraishi+2008}.

\section{Effect of Change in Damping Timescale of Planetesimals}
\label{sec:Model2}
Gap formation affects the surface density profile $\Sigmaf$ and, thereby, the rotation velocity $\vKep (1-\etadisk)$ in a circumstellar disc.
The change in surface density leads to changing the gas drag strength and the damping timescales, $\taudampaxi$, $\taudampecc$ and $\taudampinc$ almost equally.
The change in $\etadisk$ alters the gas drag direction and speed, which results in changing $\taudampaxi$ more largely than the other timescales.

We investigate these effects on the capture of planetesimals step by step in this section.
First, we describe the model for the surface density profile in the gap in \refsec{sec:Model2_method}.
In \refsec{sec:Model2_Result1}, we investigate the capture of planetesimals using the gap model, but assuming that $\etadisk$ is zero.
Then, we take into account the effect of $\etadisk$ according to the pressure gradient of the disc model in \refsec{sec:Model2_Result2}.

\subsection{Gap Model}\label{sec:Model2_method}
To take into account the presence of a gap around the protoplanet's orbit in the circumstellar disc, we use the empirical formula for the radial density profile in the gap that \citet{Kanagawa+2017} derived from their two-dimensional hydrodynamic simulations.

The width of the gap, $\DeltaGap$, inside which the surface density is smaller than a threshold value, $\Sigmath$, is given approximately as 
\begin{equation}
	\DeltaGap = \revised{\axiplanet} \left( 0.5 \frac{\Sigmath}{\Sigmazero} + 0.16 \right) {\Kp}^{1/4},
\end{equation}
where $\rplanet$ is the semi-major axis of the planetary orbit, and $\Kp$ is a non-dimensional quantity given by
\begin{equation}\label{eq:Method_Sigma_Kp}
	\Kp = \left( \frac{\Mplanet}{\Mstar} \right)^2 \left( \frac{\hscale}{\revised{\axiplanet}} \right)^{-3} \alphaPPD^{-1};
\end{equation}
$\alphaPPD$ is the viscosity factor introduced by \citet{Shakura+1973}.
The radial distribution of the surface density in the circumstellar disc with the gap is given as
\begin{equation}\label{eq:Method_Sigma}
  \Sigmaf (r) = \left\{
  \begin{array}{ll}
    \Sigmamin / \Sigmazero & \text{for $| r - \revised{\axiplanet} | < \Rone$}, \\
    \Sigmagap / \Sigmazero & \text{for $\Rone < | r - \revised{\axiplanet} | < \Rtwo$}, \\
    1 & \text{for $\Rtwo < | r - \revised{\axiplanet} |$},
  \end{array}
  \right.
\end{equation}
where
\begin{eqnarray}\label{eq:Method_Sigma_Gap}
	\frac{\Sigmagap}{\Sigmazero}  &=& 4.0 {\Kp}^{-1/4} \frac{| r - \revised{\axiplanet} |}{\revised{\axiplanet}} -0.32, \\
	\Rone &=& \left( \frac{\Sigmamin}{4 \Sigmazero} + 0.08 \right) {\Kp}^{1/4} \revised{\axiplanet}, \\
	\Rtwo &=& 0.33 {\Kp}^{1/4} \revised{\axiplanet},
\end{eqnarray} 
and $\Sigmamin$ is the surface density at the gap bottom, which is given by
\begin{equation}\label{eq:Method_Sigma_min}
	\frac{\Sigmamin}{\Sigmazero} = \frac{1}{1+ 0.04 \K}
\end{equation}
with
\begin{equation}\label{eq:Method_Sigma_K}
	\K = \left( \frac{\Mplanet}{\Mstar} \right)^{2} \left( \frac{\hscale}{\revised{\axiplanet}} \right)^{-5} {\alphaPPD}^{-1}.
\end{equation}

\begin{figure}
  \begin{center}
    \includegraphics[width=80mm]{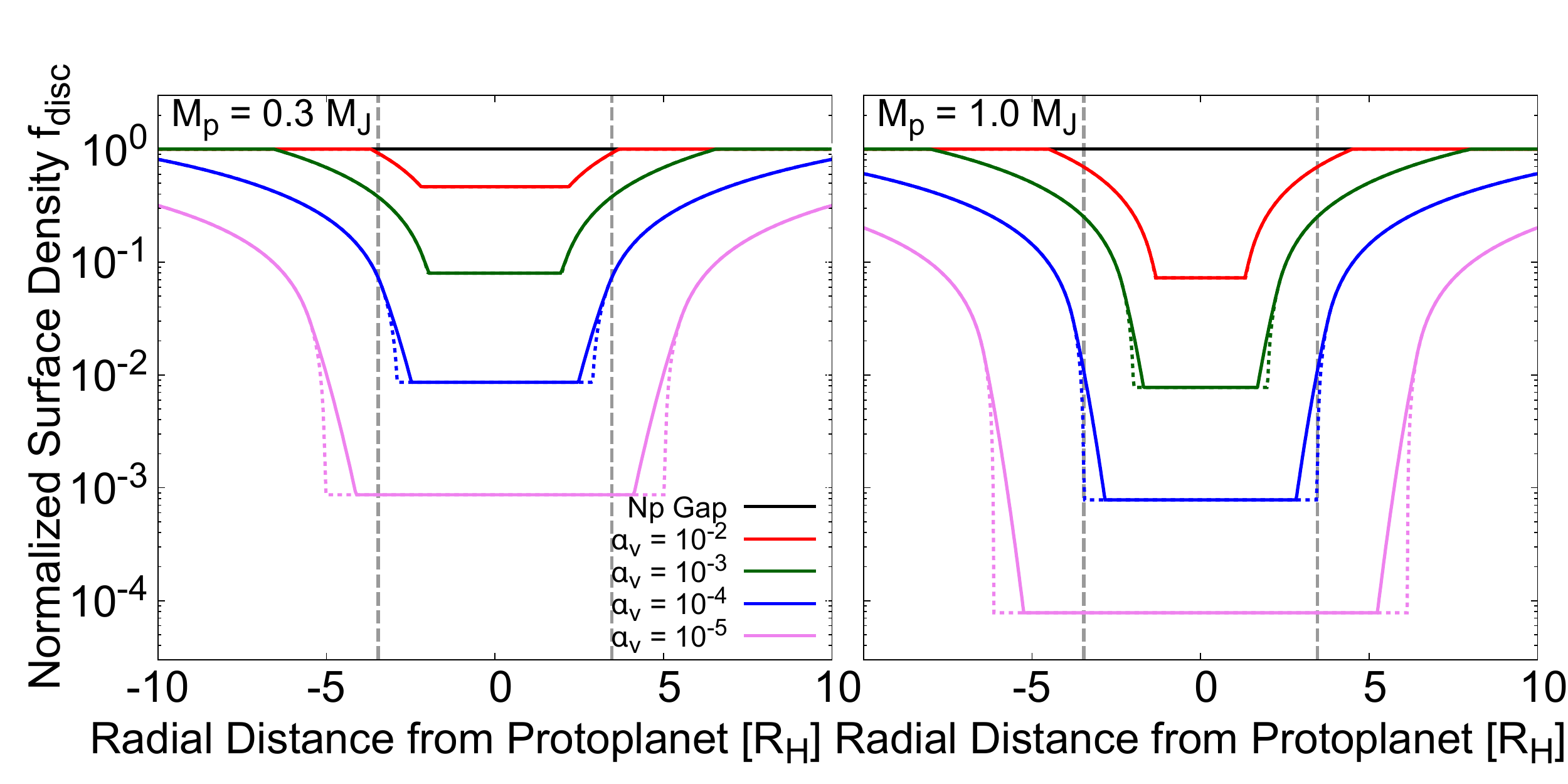}
    \caption{
    Radial profile of surface density in the circumstellar disc with a gap opened by the protoplanet, which we model modifing the empirical formula of \citet{Kanagawa+2017} using Rayleigh criterion.
    The horizontal axis is the radial distance measured from the protoplanet center in the unit of the Hill radius $\RHill \equiv ( \Mplanet / 3 \Mstar )^{1/3} \rplanet$. 
    The left and right panels show the cases of the protoplanet mass $\Mplanet$ = 0.3~$\Mjupiter$ and 1.0~$\Mjupiter$ ($\Mjupiter$: Jupiter mass), respectively.
    Different types of line are for different values of the $\alpha$-viscosity $\alphaPPD$, as indicated in the left panel.
    The dashed lines are models obtained in \citet{Kanagawa+2017}.
    The vertical dashed lines ($|\rplts-\rplanet|/\RHill$ = 2$\sqrt{3}$) indicate the outer boundary of the feeding zone when a planetesimal's eccentricity $\ecc = 0$.
    }
    \label{fig:Surface_Density_Profile_Gap}
  \end{center}
\end{figure}

For the gapped-disc models, the hydrodynamic stability of the surface density profile must be checked by the well-known Rayleigh criterion \citep[e.g.][]{Chandrasekhar+1961, Tanigawa+2007},
\begin{equation}\label{eq:Raylegih_Criterion}
	\deriv{\vgas}{x} > - 2 \Omegap,
\end{equation}
where $x = |r - \revised{\axiplanet}|$ and $\Omegap$ is the angular velocity of the protoplanet.
In the gap model given by \refeqs{eq:Method_Sigma}, the Rayleigh criterion is violated in the vicinity of the protoplanet. 
The unrealistically steep gradient yields such a large difference in velocity between gas and planetesimals that the gas drag makes a considerable effect on the dynamics of the planetesimals.
Therefore, we modify the gap structure so that the surface density profile is smoothly connected with what fulfills the Rayleigh condition marginally. 
We denote the point of connection by $\xm$ and the gas velocity at that point by $\vgasm$. 
Thus, at $x < \xm$, the gas velocity is given by
\begin{equation}\label{eq:vel_Rayleigh}
	\vgas = \vgasm - 2 \Omegap \left( x - \xm \right).
\end{equation}
The surface gas density is given by integrating the equation of motion,
\begin{equation}\label{eq:eqm_disk_gas}
	3 {\Omegap}^2 x + 2 \Omegap \vgas - {\csound}^2 \pder{\ln{\Sigmagas}}{x} = 0,
\end{equation}
inward from $x$ = $\xm$.
When the surface density thus obtained is smaller than that of the gap bottom obtained by \refeqs{eq:Method_Sigma_min}, we assume $\Sigmaf = \Sigmamin / \Sigmazero$.
Examples of the surface density profile are shown in \reffig{fig:Surface_Density_Profile_Gap}.
In this gap model, the gap width and depth depend on the protoplanet mass and disc gas viscosity.

We regard the growth rate of the protoplanet $\Mdotplanet$, the size of planetesimals $\Rplts$ and the viscosity parameter $\alphaPPD$ as free parameters.
Firstly, we focus on the strength of gas drag.
The surface density profile is given by \refeqs{eq:Method_Sigma}, but disc gas rotates around the central star with the Kepler velocity (namely, $\etadisk=0$).
We call this model the zero-$\etadisk$ model (or $\Mgwoe$). 
Secondly, we focus on the direction of gas drag.
The surface density profile is given by \refeqs{eq:Method_Sigma} and $\etadisk$ is calculated from \refeqs{eq:eta_disk}.
We call this model the nonzero-$\etadisk$ model (or $\Mgwe$).
Parameters and models are summarised in \reftab{tb:settings}.

\subsection{Effect of gas drag strength}\label{sec:Model2_Result1}
\begin{figure}
    \begin{center}
    \includegraphics[width=80mm]{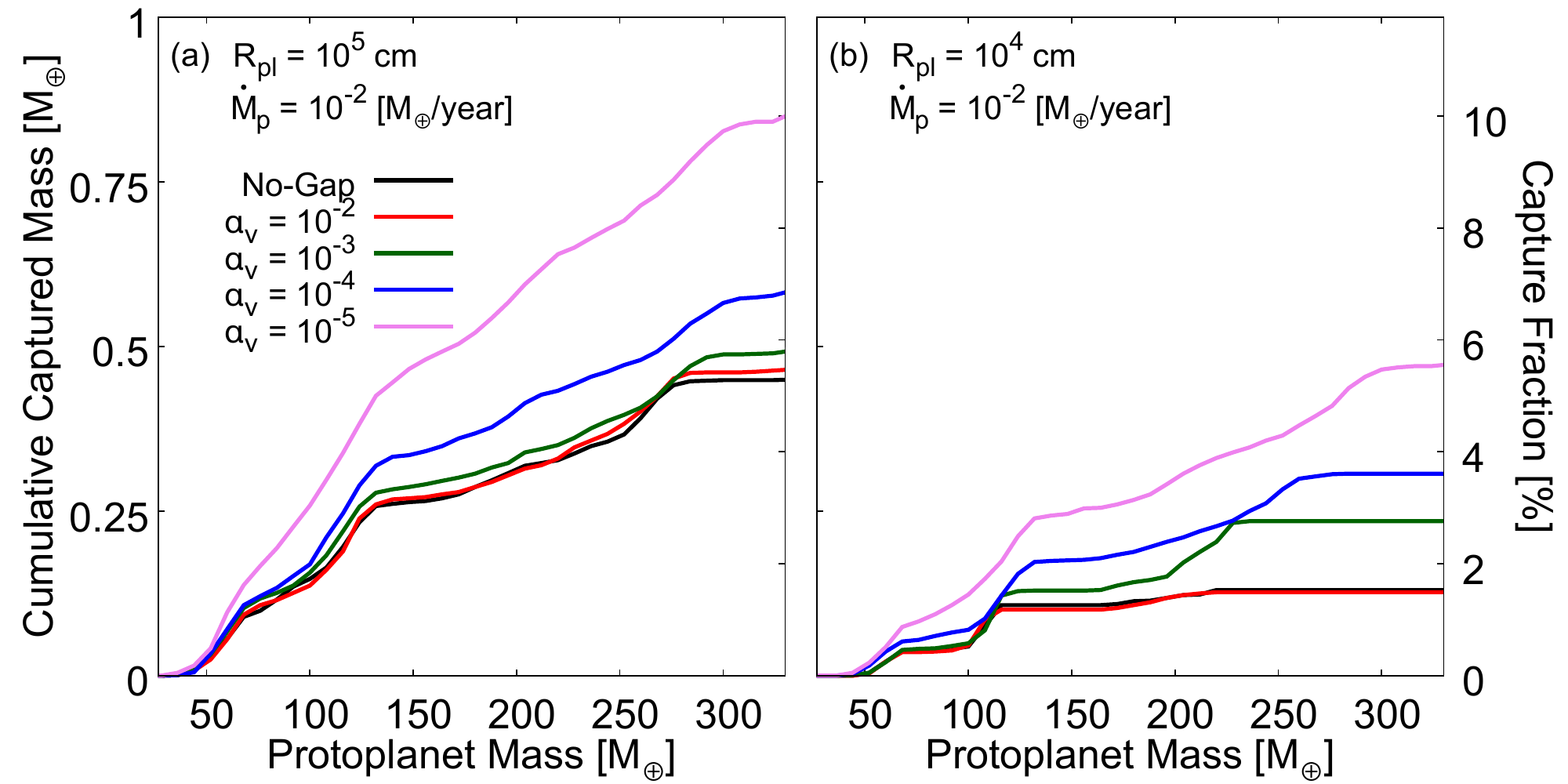}
    \caption{
    Temporal change in the cumulative mass of planetesimals captured by the protoplanet in the zero-$\etadisk$ model ($\Mgwoe$) for the protoplanetary growth rate $\Mdotplanet$ = $1 \times 10^{-2}$~$ \Mearth / \yr$.
    In panels~(a) and (b), the planetesimal radius $\Rplts$ is $1 \times 10^5$~$\cm$ and $1 \times 10^4$~$\cm$, respectively.
    The left-$y$ axis is the absolute captured mass, while the right-$y$ axis is the mass relative to the initial total mass of planetesimals (called the capture fraction).
    The $x$ axis is the protoplanet mass, which is equivalent to time.
    The lines are colour-coded according to $\alpha$-viscosity $\alphaPPD$, as indicated in the panel, except for the black line which represents the case of the unperturbed disc ($\Mupd$).
    }
    \label{fig:Results_Gap_1}
    \end{center}
\end{figure}


Figure~\ref{fig:Results_Gap_1} shows the cumulative captured mass as a function of the protoplanet mass for $\Mdotplanet$ = $1 \times 10^{-2}$~$\Mearth {\yr}^{-1}$ obtained in the case of the gapped disc with Kepler rotation (i.e., $\etadisk = 0$; $\Mgwoe$). 
In panels~(a) and (b), $\Rplts$ = $1 \times 10^5$ and $1 \times 10^4$~$\cm$, respectively. 
The lines are colour-coded according to the value of $\alphaPPD$ except for the black one representing the no-gap case.

In both panels, because of reduced gas drag due to gap formation, the cumulative captured mass becomes larger than in the unperturbed (no-gap) disc and larger with decreasing $\alphaPPD$. 
Such an effect turns out to be noticeable for $\alphaPPD \lesssim 10^{-3}$. 
This is because the gap becomes wider than the feeding zone for $\alphaPPD \lesssim 10^{-3}$, as found in \reffig{fig:Surface_Density_Profile_Gap}.
In the small-$\Rplts$ case shown in panel~(b), where the damping timescale is shorter than in the high-$\Rplts$ case, a larger number of planetesimals are eliminated from the feeding zone, reducing the cumulative captured mass as a whole.

\begin{figure}
    \centering
    \includegraphics[width=80mm]{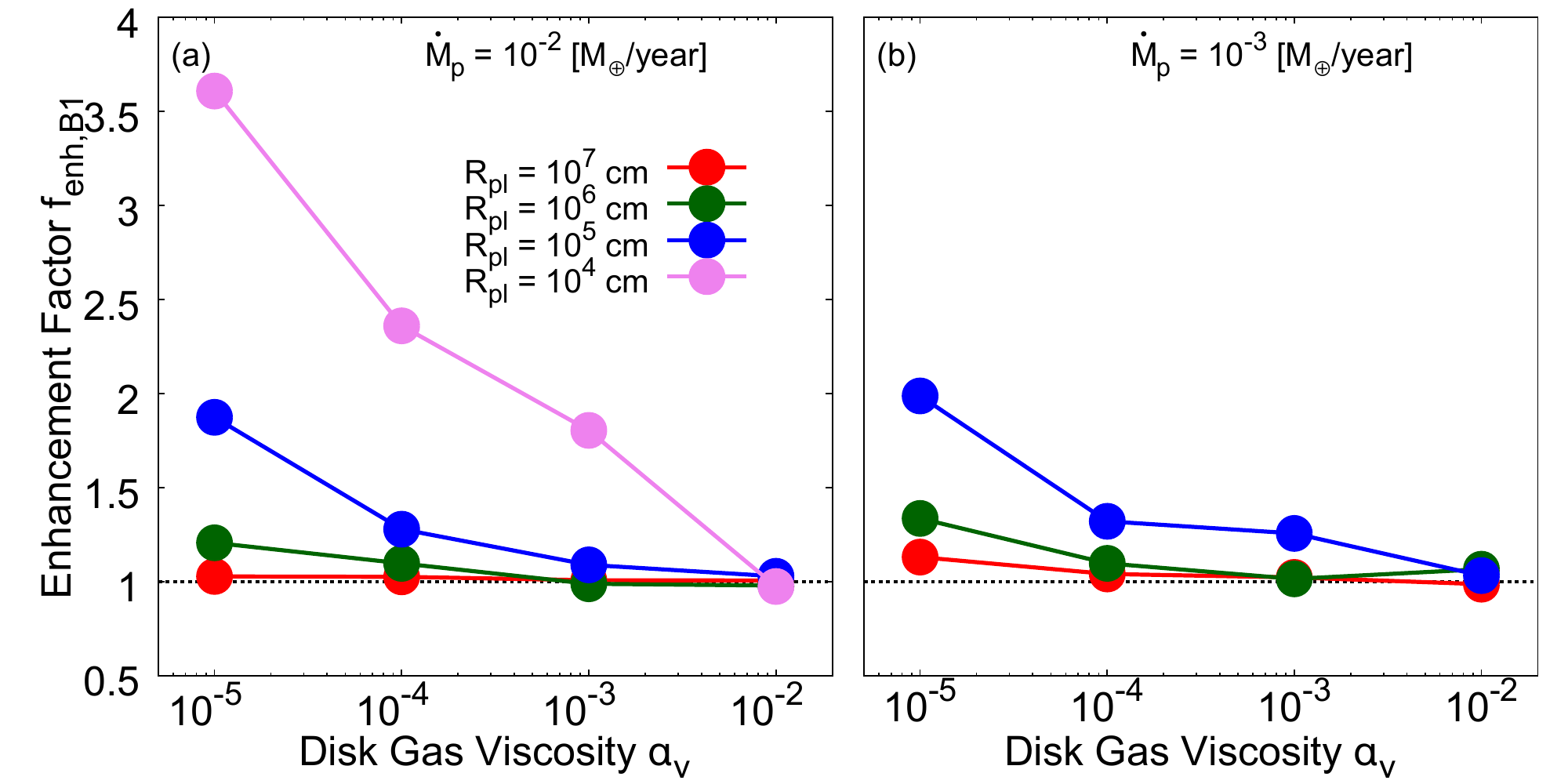}
    \caption{
    The total mass of captured planetesimals in the zero-$\etadisk$ model ($\Mgwoe$) relative to that in the unperturbed disk ($\Mupd$) (the enhancement factor, $\fGwoE$; see \refeqs{eq:define_feff}) versus the $\alpha$-viscosity $\alphaPPD$ for $\Mdotplanet$ = $1 \times 10^{-2} \Mearth {\yr}^{-1}$ (left panel)
    and $1 \times 10^{-3} \Mearth {\yr}^{-1}$ (right panel).
    The red, green, blue and magenta lines show the results for the planetesimal radius $\Rplts$ = $1 \times 10^{7}$, $1~\times~10^{6}$, $1~\times~10^{5}$, and $1~\times 10^{4}$~$\cm$, respectively. 
    The dotted line indicates $\fGwoE=1$ for an eye guide.
    In the right panel, the result for $\Rplts = 10^4 \cm$ is not defined because no planetesimals are captured in model A.
    }
    \label{fig:Enh_mul_eta0}
\end{figure}

For a clear understanding of the effect of gap formation on the capture of planetesimals, we introduce an enhancement factor defined as
\begin{equation}\label{eq:define_feff}
   \fGwoE \equiv \frac{\McapGwoE}{\McapUD},
\end{equation}
where $\McapUD$ and $\McapGwoE$ are the total captured mass in $\Mupd$ and $\Mgwoe$, respectively.
Figure~\ref{fig:Enh_mul_eta0} shows the calculated relationship between $\fGwoE$ and $\alphaPPD$ for two different values of $\Mdotplanet$ and four different values of $\Rplts$. 
It is found that $\fGwoE$ ranges between 1.0 and 3.6, depending on $\alphaPPD$, $\Rplts$ and $\Mdotplanet$.
Note that for $\Mdotplanet$ = $1 \times 10^{-3}~\Mearth~{\yr}^{-1}$ and $\Rplts$ = $1 \times 10^{4}~\cm$, $\fGwoE$ is not defined because $\McapUD=0$.

The enhancement factor $\fGwoE$ increases with decreasing $\alphaPPD$ for given $\Rplts$ and $\Mdotplanet$. 
Again, this is because the gap becomes wider \revised{and deeper} with decreasing $\alphaPPD$, so that eccentricity damping and elimination of planetesimals become less effective.
Also, $\fGwoE$ becomes larger with decreasing $\Rplts$ except for the cases of $\revised{\alphaPPD~=}~1~\times 10^{-2}$.
When $\Rplts$ is as large as $1 \times 10^7 \cm$, the timescale of eccentricity damping is much longer than the protoplanet's growth timescale, so that the elimination of planetesimals from the feeding zone is ineffective even in the unperturbed disc.
Thus, the effect of the gap is negligibly small and therefore $\fGwoE = 1$.
However, for $\Rplts \leq 1 \times 10^6 \cm$, eccentricity damping occurs efficiently enough that a significant fraction of planetesimals are pushed away from the feeding zone.
The effect of the gap reduces such an elimination effect and thus $\fGwoE$ is larger for smaller $\Rplts$.

\subsection{Effect of gas flow velocity}\label{sec:Model2_Result2}
\begin{figure}
    \begin{center}
    \includegraphics[width=80mm]{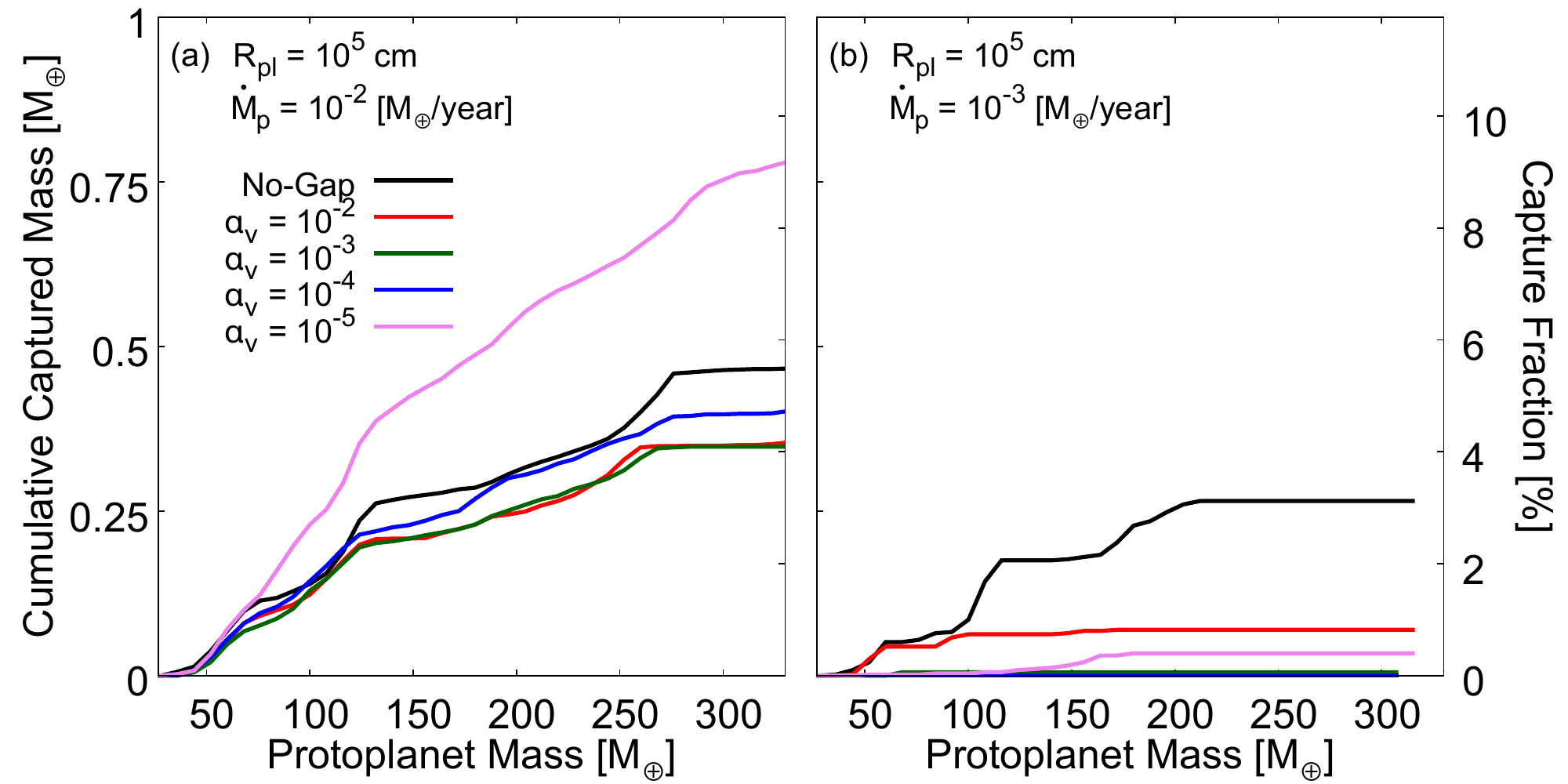}
    \caption{
    Same as \reffig{fig:Results_Gap_1}, but for the nonzero-$\etadisk$ model ($\Mgwe$). 
    The planetesimal radius $\Rplts$ is $1 \times 10^5~\cm$. 
    In the left and right panels, the protoplanet's growth rate $\Mdotplanet$ is $1 \times 10^{-2}~\Mearth~{\yr}^{-1}$ and $1 \times 10^{-3}~\Mearth~{\yr}^{-1}$, respectively.
    }
    \label{fig:Results_Gap_2}
    \end{center}
\end{figure}
Figure~\ref{fig:Results_Gap_2} is the same as \reffig{fig:Results_Gap_1}, but for the nonzero-$\etadisk$ model ($\Mgwe$).
In this case, unlike in \reffig{fig:Results_Gap_1}, the cumulative captured mass is smaller in the gapped disc (coloured lines) than in the unperturbed disc  (black line), except for the case of $\alphaPPD$ = $1 \times 10^{-5}$ (magenta line).
On the gap slopes, $|\etadisk|$ takes large values, so that planetesimals drift rapidly in the directions away from the protoplanet,  resulting in reduction in the number of planetesimals that enter the feeding zone and thereby in the capture rate of planetesimals.
Such an effect appears basically regardless of $\alphaPPD$, because the gap extends beyond the feeding zone (see \reffig{fig:Surface_Density_Profile_Gap}) and $|\etadisk|$ is sufficiently large even near the edges of the gap slopes in the nonzero-$\etadisk$ models. 
The reason why the cumulative captured mass for $\alphaPPD = 1 \times 10^{-5}$ and $\Mdotplanet = 1 \times 10^{-2} \Mearth \, \yr^{-1}$ (magenta line) is larger than that for the unperturbed disc case (black line) is that the effect of reduction in gas drag strength, which was shown in section~\ref{sec:Model2_Result1}, dominates over that of change in pressure gradient.
Thus, compared to the unperturbed disc model, the total captured mass is larger (smaller) when the effect of gas drag strength is more (less) effective than that of $\etadisk$.

\begin{figure}
    \centering
    \includegraphics[width=80mm]{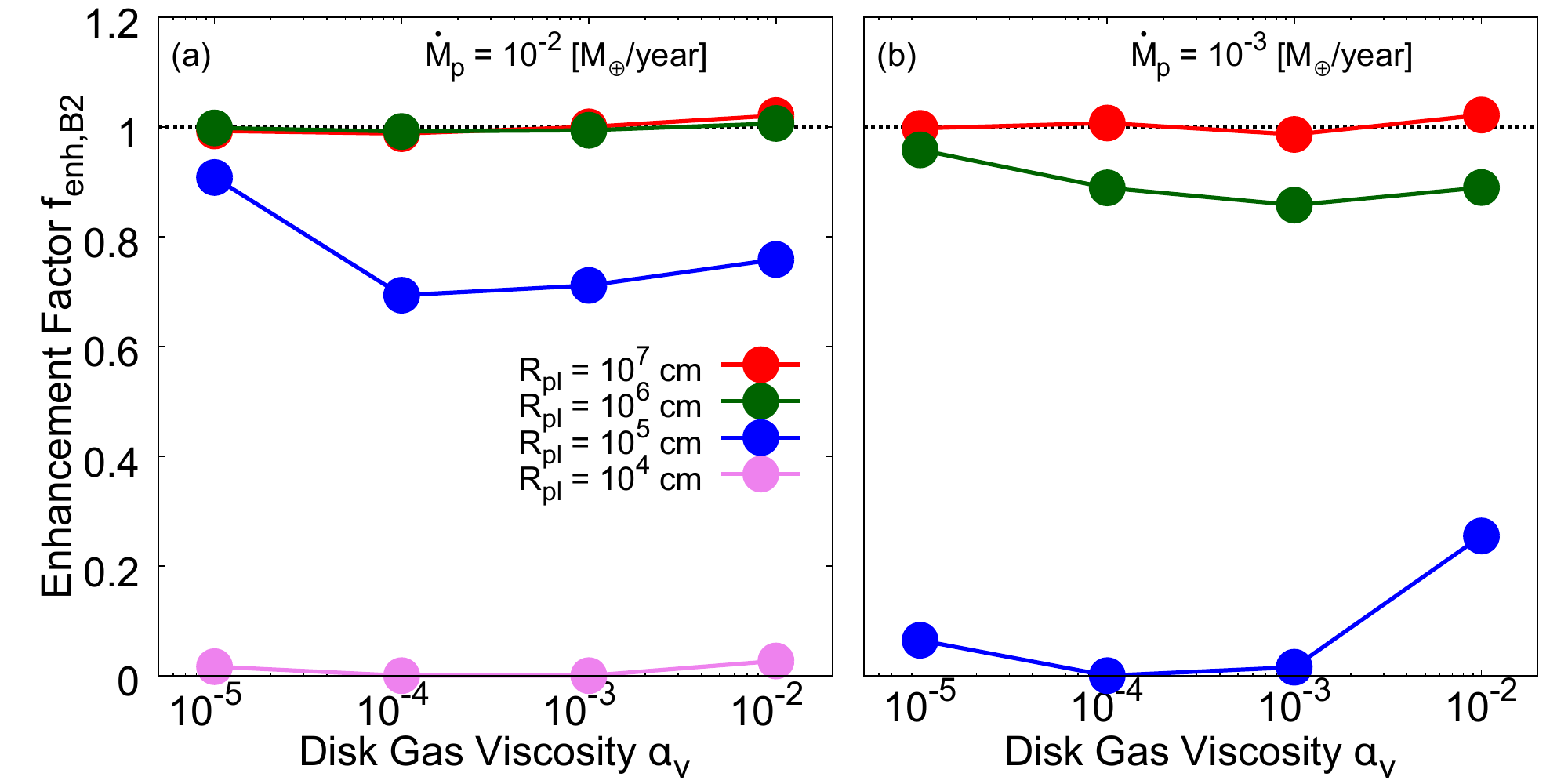}
    \caption{
    The total mass of captured planetesimals in the nonzero-$\etadisk$ model ($\Mgwe$) relative to that in the zero-$\etadisk$ model ($\Mgwoe$) (the enhancement factor, $\fGwE$; see \refeqs{eq:define_fenh2}) versus the $\alpha$-viscosity $\alphaPPD$ 
    for $\Mdotplanet$ = $1 \times 10^{-2} \Mearth {\yr}^{-1}$ (left panel) and $\Mdotplanet$ = $1 \times 10^{-3} \Mearth {\yr}^{-1}$ (right panel).
    The red, green, blue and magenta lines show the results for $\Rplts$ = $1~\times~10^{7}$, $1~\times~10^{6}$, $1~\times~10^{5}$, and $1~\times 10^{4}$~$\cm$, respectively. 
    The dotted line indicates $\fGwE=1$ for an eye guide.
    }
    \label{fig:Enh_mul_KR}
\end{figure}

For a clear understanding of the effect of $\etadisk$ on the capture of planetesimals, we introduce an enhancement factor for $\Mgwe$ defined as
\begin{equation}\label{eq:define_fenh2}
   \fGwE \equiv \frac{\McapGwE}{\McapGwoE},
\end{equation}
where $\McapGwE$ is the total mass of captured planetesimals in $\Mgwe$.
Figure~\ref{fig:Enh_mul_KR} shows $\fGwE$ versus $\alphaPPD$ for two different values of $\Mdotplanet$ and four different values of $\Rplts$.
The results are divided into three groups: 
In the first group including the cases of ($\Rplts$ [$\cm$], $\Mdotplanet$ [$\Mearth {\yr}^{-1}$]) = ($1 \times 10^7$, $1 \times 10^{-2}$), ($1 \times 10^6$, $1 \times 10^{-2}$) and ($1 \times 10^7$, $1 \times 10^{-3}$), $\fGwE$ is almost unity.
This is because the timescale of change in semi-major axis $\taudampaxi$ is much longer than the growth timescale of the protoplanet $\tauacc$, meaning that the radial drift  on the gap slopes is too slow to affect the cumulative captured mass $\Mcap$.
In the second group including the cases of $(\Rplts, \Mdotplanet) = (1 \times 10^4, 1 \times 10^{-2})$ and $(1 \times 10^5, 1 \times 10^{-3})$, $\fGwE$ is significantly smaller than unity because $\taudampaxi \ll \tauacc$ on the gap slopes.
In the last group including the cases of $(\Rplts, \Mdotplanet) = (1 \times 10^5, 1 \times 10^{-2})$ and $(1 \times 10^6, 1 \times 10^{-3})$, $\taudampaxi \sim \tauacc$ on the gap slopes and thus $\fGwE$ takes intermediate values between zero and unity.

\subsection{Change of Jacobi Energy}\label{sec:Results_CJ}
As indicated in \refeqs{eq:accretion_rate2}, the capture rate of planetesimals $\Mdotcap$ is proportional to the effective surface density of planetesimals inside the feeding zone $\Sigmaeff$ (see \refsec{sec:Result_1_orb} for its definition), which depends on the Jacobi energy $\Ejacobi$.
Here, we give an interpretation to numerical results obtained in Sections~\ref{sec:Model2_Result1} and \ref{sec:Model2_Result2}, in terms of the Jacobi energy explained in \refsec{sec:Result_1_orb}.

\begin{figure}
    \begin{center}
    \includegraphics[width=80mm]{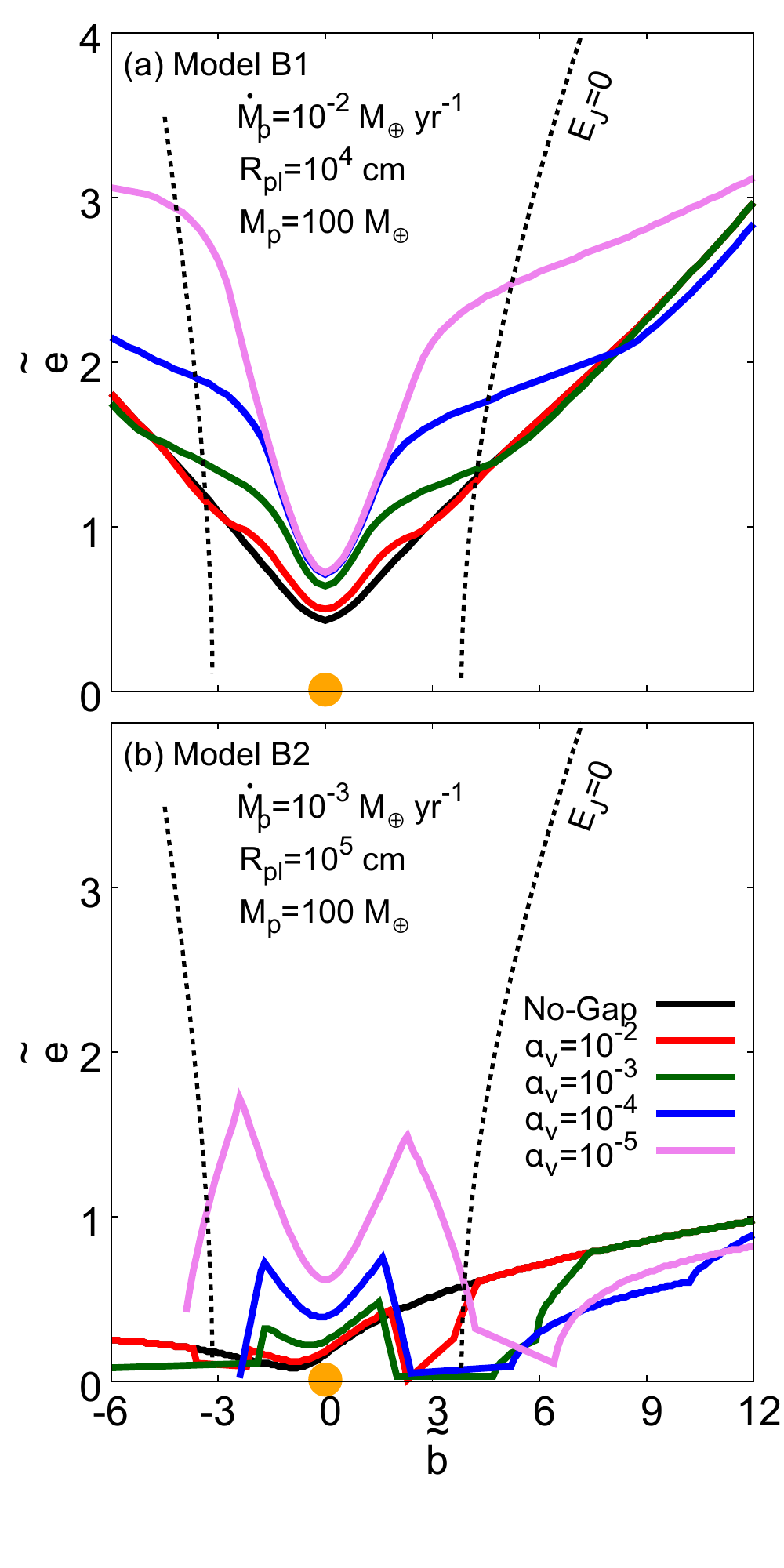}
    \caption{
    The effect of the disc gas viscosity ($\alphaPPD$; see \refeqs{eq:Method_Sigma_Kp}) on the change in the Jacobi energy $\Ejacobi$ (a) for the zero-$\etadisk$ model ($\Mgwoe$) and (b) for the nonzero-$\etadisk$ model ($\Mgwe$).
    The magenta, blue, green, and red lines represent $\Ejdotnrm=0$ for $\alphaPPD$ = $1 \times 10^{-5}$, $1 \times 10^{-4}$, $1 \times 10^{-3}$ and $1 \times 10^{-2}$, respectively, on the plane of the normalised semi-major axis $\bnrm$ vs. normalised eccentricity $\eccnrm$. 
    The protoplanet mass $\Mplanet$ is $100 \Mearth$.
    In panel (a), the protoplanet growth rate $\Mdotplanet$ is $ 1 \times 10^{-2}~\Mearth~{\yr}^{-1}$ and the planetesimal radius $\Rplts$ is $1 \times 10^{4}~\cm$.
    In panel (b), $\Mdotplanet$ = $ 1 \times 10^{-3}~\Mearth~{\yr}^{-1}$ and $\Rplts$ = $1 \times 10^{5}~\cm$.
    The black solid line shows the case of the unperturbed disc model ($\Mupd$).
    The black dotted line shows the boundary of the feeding zone (i.e., $\Ejacobi = 0$).
    }
    \label{fig:ResultsANL_alpha_1}
    \end{center}
\end{figure}

First, we focus on the effect of decrease in gas drag strength.
Figure~\ref{fig:ResultsANL_alpha_1} shows the lines of $\Ejdotnrm=0$, below which $\Ejdotnrm > 0$ (also see Fig.~\ref{fig:ResultsORB_be}c), for different choices of $\alphaPPD$.
\revised{
Values of $\Ejdotnrm$ are calculated by numerically integrating \refeqs{eq:der_E_Jacobi_orb} with $\inc=0$ and variable $C_D$ for an orbital period.
}
Panel (a) represents the zero-$\etadisk$ disc models ($\Mgwoe$) with $\Mplanet = 100 \Mearth$, $\Mdotplanet$ = $1 \times 10^{-2}~\Mearth~{\yr}^{-1}$ and $\Rplts = 1 \times 10^{4}~\cm$; the black solid line represents the unperturbed disc model.
In every case, planetesimals with low $\eccnrm$ ($\lesssim 1$) have positive values of $\Ejdotnrm$ outside the feeding zone, which means that they take a similar evolution path shown in Fig.~\ref{fig:ResultsORB_be}c until they enter the feeding zone.
After entering the feeding zone, however, planetesimals take different paths depending on the gap size.
As illustrated in Fig.~\ref{fig:ResultsANL_alpha_1}a, the area of $\Ejdotnrm>0$ expands with decreasing $\alphaPPD$ inside the feeding zone.
Because of the expansion of the area of $\Ejdotnrm>0$ and $\Ejacobi > 0$, the planetesimals can stay inside the feeding zone for a longer time and $\Sigmaeff$ becomes higher in a deeper and wider gap.

Second, we focus on the effect of the change in gas flow direction and velocity.
Panel (b) of \reffig{fig:ResultsANL_alpha_1} is the same as panel~(a), but for the nonzero-$\etadisk$ disc model ($\Mgwe$). 
The values of $\Mdotplanet$ and $\Rplts$ adopted here are  different from those in panel~(a) and are $1 \times 10^{-3}~\Mearth~{\yr}^{-1}$ and $1 \times 10^{5}~{\cm}$, respectively.
The change of gas flow inside the gap turns out to make a great change to the line of $\Ejdotnrm=0$.
In the three cases of $\alphaPPD \leq 1 \times 10^{-3}$ (green, blue and magenta lines), planetesimals are unable to reach the feeding zone, because the evolution path is truncated by the line of $\Ejdotnrm = 0$ (solid line) before it reaches the line of $\Ejacobi = 0$ (dashed line).
Even when the domain of $\Ejdotnrm>0$ expands in the feeding zone, $\Sigmaeff$ never increases due to the prohibited supply of planetesimals.
In the case of $\alphaPPD = 1 \times 10^{-2}$, planetesimals barely reach the feeding zone when $\Mplanet \lesssim 100 \Mearth$, but are never captured when $\Mplanet \gtrsim 100 \Mearth$.

\section{Effect of Change in Growth Timescale of Protoplanet}
\label{sec:Model3}
In the previous sections, we have assumed a constant growth rate of the protoplanet. 
In reality, however, the protoplanet grows at a time-dependent rate, which depends on the protoplanet's mass and surrounding gas density, namely the gap structure.
Here, adopting a simple model of protoplanetary growth via gas accretion, we demonstrate that the capture of planetesimals is affected by growth modes of the protoplanet.

\subsection{Model of Protoplanet's Growth via Gas Accretion}\label{sec:Mod3_Growth_Model}
We consider protoplanet's growth via gas accretion after the onset of the runaway gas accretion. 
In the first phase, the protoplanet's growth is controlled by the Kelvin-Helmholtz contraction of the gaseous envelope \citep[][]{Bodenheimer+1986,Tajima+1997,Ikoma+2000}.
We adopt the fitting formula of the growth rate numerically derived by \citet{Tajima+1997},
\begin{equation}\label{eq:ACC_Tajima}
  \MdotplanetR =1.4 \times 10^{-11} \left( \frac{\Mplanet}{1 \Mearth} \right)^{4.9}  {\rm \Mearth {\yr}^{-1}},
\end{equation}
which is valid for $\Mplanet \gtrsim$ 30~$\Mearth$. 
\revised{Since the growth rate increases rapidly with increasing mass of the protoplanet, it exceeds the maximum supply rate of disc gas at a certain mass} \citep[][]{Tanigawa+2002}.
Thus, the second phase is characterised by supply-limited gas accretion and called the supply-limited phase.

In the second phase, we adopt the model of the supply-limited gas accretion formulated by \citet{Tanigawa+2007}.
Based on the numerical results of \citet{Tanigawa+2002}, which demonstrated that disc gas flows into the protoplanet's Roche lobe through the two bands around $\xgap \equiv \revised{|r - \axiplanet|} \sim 2 \RHill$, \citet{Tanigawa+2007} derived the maximum supply of disc gas as
\begin{equation}\label{eq:ACC_SLP}
  \MdotplanetS= \dot{A} \Sigma_\mathrm{band},
\end{equation}
where
\begin{equation}\label{eq:ACC_SLP_Area}
  \dot{A} \simeq 0.29 \left( \frac{\hscale}{\revised{\axiplanet}} \right)^{-2} \left( \frac{\Mplanet}{\Mstar} \right)^{4/3} \rplanet^2 \Omegap
\end{equation}
and $\Sigma_\mathrm{band}$ is the surface density of disc gas at $\xgap = 2 \RHill$.
The surface density $\Sigma_\mathrm{band}$ is affected not only by gap formation, but also global disc evolution, in particular, disc dispersal.
Since the detailed treatment of global disc evolution is beyond the scope of this study, we assume that the surface gas density outside the gap $\Sigmazero$ is given simply as
\begin{equation}\label{eq:Discussion_disc_Disp}
   \Sigmazero = \Sigmazeroint \exp{\left(- \frac{t - \tcont}{\taupep} \right)},
\end{equation}
where $\Sigmazeroint$ is the initial, unperturbed disc surface density, $\taupep$ is the timescale of disc dissipation, and $\tcont$ is a parameter that determines the onset of disc dissipation.
The surface density in the accretion band $\Sigma_\mathrm{band}$ is given by combination of Eqs.~(\ref{eq:Method_Sigma_Gap}) and (\ref{eq:Discussion_disc_Disp}), which is applied to Eq.~(\ref{eq:ACC_SLP}) to obtain the maximum supply rate. 
In summary, the protoplanet's growth rate is given as the smaller of $\MdotplanetR$ and $\MdotplanetS$, namely
\begin{equation}\label{eq:ACC_Rate}
  \Mdotplanet = {\rm min} \left( \MdotplanetR , \MdotplanetS \right).
\end{equation}

\begin{figure}
    \begin{center}
    \includegraphics[width=80mm]{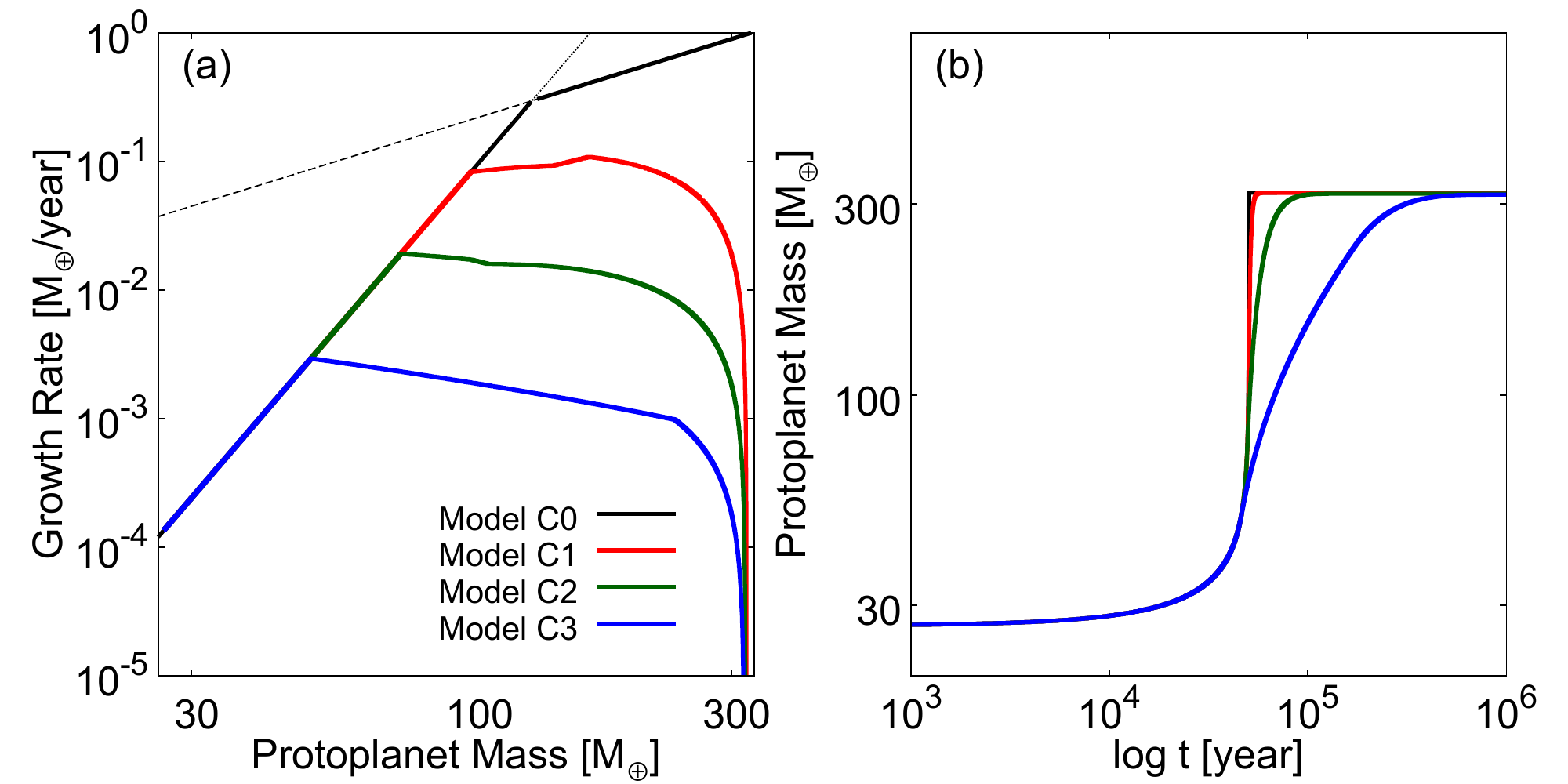}
    \caption{
    Models of the protoplanet's growth via gas accretion. 
    See Section~\ref{sec:Mod3_Growth_Model} for the details. 
    Panel (a) shows the growth rate as a function of the protoplanet mass, while panel (b) shows the protoplanet mass as a function of time for three different choices of the $\alpha$-viscosity $\alphaPPD$ and the disc dispersal timescale $\taupep$,
    \revised{
    $\alphaPPD=10^{-2}$ and $\taupep=10^3~\yr$ for model C1 (red line), $\alphaPPD=10^{-3}$ and $\taupep=10^4~\yr$ for model C2 (green line) and
    $\alphaPPD=10^{-4}$ and $\taupep=10^5~\yr$ for model C3 (blue line).}
    The case without gap formation nor disc dispersal \revised{(model C0)} is represented by the black line.
    }
    \label{fig:Discussion_Mp_Mdot}
    \end{center}
\end{figure}

Figure~\ref{fig:Discussion_Mp_Mdot} shows (a) the protoplanet's growth rate and (b) the protoplanet's mass thus calculated for three different choices of $\alphaPPD$ and $\taupep$ as indicated in each panel.
The value of $\tcont$ is chosen so that the protoplanet mass is 1~$\Mjupiter$ at the end of integration.
The black solid lines represent the case without gap formation nor disc dispersal (i.e., $\alphaPPD \rightarrow \infty$ and $\taupep \rightarrow \infty$), which \citet{Shiraishi+2008} assumed.
The two parameters control the protoplanet mass at which the transition from the contraction-limited and supply-limited growth occurs.
The smaller the viscosity $\alphaPPD$, the smaller the transition mass is, because the gap is wider, reducing the gas surface density in the accretion band.
In the early supply-limited growth phase, the growth rate decreases gradually, but drops rapidly in the late phase, because of disc dispersal.
In every case, the growth rate in the late phase is smaller by several orders of magnitude than that in the case without gap formation nor disc dispersal.
Such a low growth rate is expected to reduce the capture rate of planetesimals, as seen above.
On the other hand, the phase of low growth rate lasts long, as shown in the Fig.~\ref{fig:Discussion_Mp_Mdot}b. 
The growth models are labelled in the order of growth speed as \ModelCa-C3, namely, \ModelCa~(black) is the fastest model and \ModelCd~(blue) is the slowest model. 
The parameters used in those models are listed in \reftab{tb:settings}.
Below we investigate such competing effects on the total captured mass of planetesimals.

\subsection{Calculation of capture of planetesimals}\label{sec:Mod3_Result}

\begin{figure}
  \begin{center}
    \includegraphics[width=80mm]{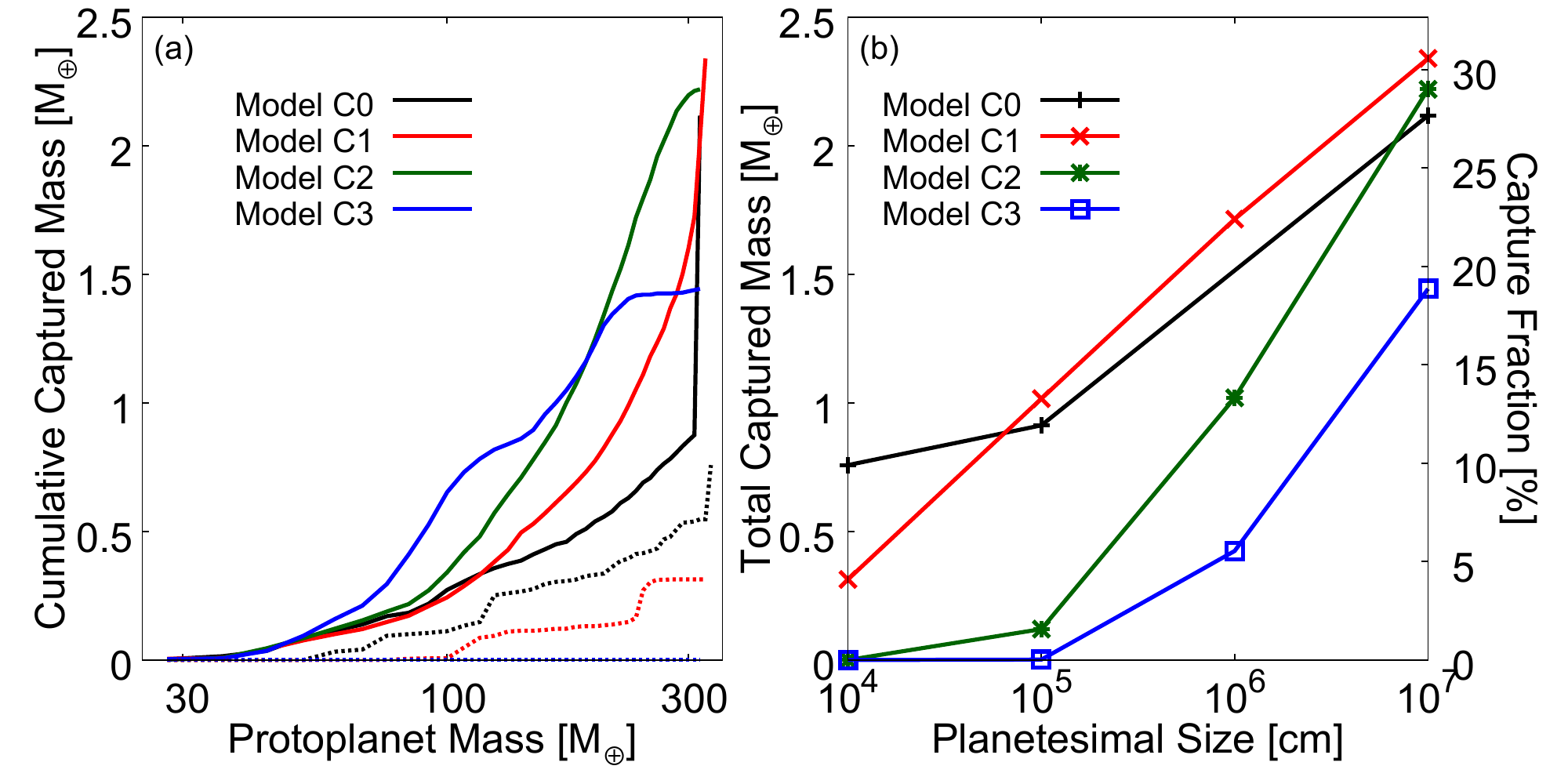}
    \caption{
    (a) The cumulative mass of planetesimals captured by the protoplanet that grows at the rate modelled in Section~\ref{sec:Mod3_Growth_Model}. 
    The numerical settings are given in \reftab{tb:settings}. 
    The lines are colour-coded according to the growth model as also indicated in each panel. 
    The solid and dotted lines show the cases of $\Rplts~=~1~\times~10^7~\cm$ and $1~\times 10^4~\cm$, respectively.
    (b) The total captured mass of planetesimals with the size of planetesimals.
    The lines are colour-coded according to the growth model as also indicated in each panel. 
    }
    \label{fig:DiscussionTot_Mcap}
  \end{center}
\end{figure}


Here we perform orbital integration of planetesimals around the protoplanet growing at a rate modelled above.  
The integration is continued until the surface density becomes low enough that the protoplanet mass increases no more, namely, $\tauacc > 1 \times 10^7$~year ($\gg \taupep$). 
The initial protoplanet mass $\Mpint$ is assumed to be the same value adopted in the previous sections.
The calculated cumulative mass of captured planetesimals is shown in \reffig{fig:DiscussionTot_Mcap}a for four different sets of $\alphaPPD$ and $\taupep$ (including $\alphaPPD \rightarrow \infty$ and $\taupep \rightarrow \infty$; i.e., no gap and disc dispersal) and two different choices of $\Rplts$.
In \reffig{fig:DiscussionTot_Mcap}b, results are summarised in the total captured mass $\Mcaptot$ with the size of the planetesimals $\Rplts$.

Firstly, we focus on $\Mcaptot$.
For the cases of $\Rplts=10^7\cm$, the fastest growth mode (C1; red) yields the largest total captured mass ($\Mcaptot \sim 2.3 \Mearth$).
As shown in Section~\ref{sec:Model2}, the change in the damping timescale due to the gap formation hardly affects the dynamics of such large-sized planetesimals.
$\Mcaptot$ takes a similar value in models~C0 (black) and C2 (green), but smaller in model C3 (blue).
Thus, it follows that this difference in $\Mcaptot$ comes from the change in the growth timescale, namely the difference of the growth model.
For the cases of $\Rplts=10^4\cm$, the changes in both timescales result in smaller $\Mcaptot$. 
In models C2 an C3, no small planetesimals are captured by the protoplanet.

Secondly, the slower the protoplanet growth is, the more rapidly the cumulative captured mass increases with the protoplanet mass \revised{in the early stage of planetary growth}.
\revised{This is because a given increase in protoplanet mass takes a longer time in the case of slower growth, so that more conjunctions and thus collisions between the planetesimals and protoplanet occur at a given protoplanet mass.}
For example, at $\Mplanet=100~\Mearth$, the slowest growth model (C3; blue) yields the largest value of $\Mcap$ among other models.
Such a difference in timing of planetesimal bombardment may affect the distribution of heavy elements inside gas giant planets.

\section{Other Effects}
\label{sec:Discussion}
\subsection{Capture of Planetesimals before Gap Formation}\label{sec:Discussion_Mint}
In this study, we have started the calculations at $\Mplanet \sim 30 \Mearth$ to focus on the effect of gap formation, ignoring the amount of planetesimals that the protoplanet captures until $\Mplanet \sim 30 \Mearth$.
Here we make a simple estimate on how much planetesimals should be captured until $\Mplanet \sim 30 \Mearth$.

The total mass of planetesimals initially existing inside the feeding zone of a protoplanet of $30~\Mearth$ at $5.2~\AU$ in MMSN is $\sim~3.3~\Mearth$ \citep{Hayashi1981}.
According to numerical simulations of the runaway gas accretion via the Kelvin-Helmholtz contraction of the gaseous envelope \citep[e.g.,][]{Ikoma+2000}, the protoplanet growth timescale is longer than $10^{5}~\yr$ just after the onset of the runaway gas accretion.
Such a growth timescale is longer than that in the \Mupd~with $\Mdotplanet=10^{-3}~\Mearth~{\yr}^{-1}$.
According to our numerical results shown in \reffig{fig:ModelA_Mcap}, even in the case of large planetesimals such as $\Rplts = 1 \times 10^7~\cm$, the fraction of captured planetesimals is smaller than $0.20$ and, thus, the cumulative captured mass of planetesimals at $\Mplanet=30\Mearth$ is $\sim 0.66$~$\Mearth$ at most.
Also in \citet{Shiraishi+2008}, the numerical results show that the cumulative captured mass at $\Mplanet = 30~\Mearth$ is smaller than $1~\Mearth$.
Hence, a protoplanet is estimated to capture a greater amount of planetesimals after gap formation than it does until that.
Thus, we conclude that the capture of planetesimals after gap formation is important to investigate the origin of the heavy elements observed inside gas giant planets.

\subsection{Effective Capture Radius of the Growing Protoplanet}\label{sec:Discussion_Rcapture}

In all the simulations above, we assumed a constant mean density of the protoplanet $\rhoave = 0.125~\g~/~{\cm}^3$, with which we calculated the protoplanet's radius  $\Rplanet$.
During the formation stage, however, the protoplanet's mean density is thought to be much lower and variable, because continuous gas accretion brings a large amount of entropy, making the protoplanet's envelope greatly inflated.
In addition, in the supply-limited growth stage, a circum-planetary disc is formed around the accreting protoplanet \citep[e.g.,][]{Tanigawa+2002}.
Both the expanding envelope and circumplanetary disc reduce the kinetic energy of incoming planetesimals and thereby enhance the capture probability of planetesimals.
Here, we make a brief discussion on such an effect.

The interaction between planetesimals and protoplanet's atmosphere before the onset of runaway gas accretion is well studied \citep[e.g.,][]{Podolak+1988,Inaba+2003,Valletta+2019}.
According to these studies, the gas drag from the atmosphere enhances the capture probability greatly.
A capture radius of a protoplanet, inside which planetesimals lose parts of their kinetic energy enough to stay in the envelope, is ten times or more larger than the radius of its solid core.
During the runaway gas accretion stage, the capture radius would be more enhanced by the inflated envelope.
Although the capture radius enhanced by the circumplanetary disc is smaller than that of the inflated envelope in runaway gas accretion phase, it is several times larger than that without circumplanetary disc \citep[e.g.,][]{Fujita+2013,Tanigawa+2014}.
In both cases, however, it is easy to say that the capture radius is much larger than the protoplanet radius we used in this study.
Detailed investigation of the effects of such enhanced capture radius is one of our future studies.

\section{Summary and Conclusion}
\label{sec:Summary}
Gas giant planets including Jupiter, Saturn, and some hot/warm Jupiters are known to contain much larger amounts of heavy elements relative to their host stars.
Such heavy elements were likely brought by planetesimals in late stages of the growth of the proto-gas giants. 
In this study, performing numerical simulations of the dynamics of planetesimals around a protoplanet growing in mass via gas accretion, we have investigated the efficiency for planetesimals to be captured by the protoplanet. 

Our main focus has been on the effects of the presence of a gap near the protoplanet's orbit in the circumstellar gaseous disc, which previous studies ignored.
Gap formation leads to reducing the disc surface density, enhancing the velocity shear between the planetesimals and disc gas by changing the gradients of surface density (or pressure), and slowing the protoplanet's growth, the effects of which we have investigated step by step in detail.
\begin{enumerate}
    \item \revised{The reduction in the disc surface density} makes damping of planetesimals' random velocities less efficient, which leads to enhancing the capture rate of planetesimals;
    \item \revised{The enhancement of the velocity shear between the planetesimals and disc gas} accelerates radial drift of planetesimals, which leads to reducing the capture rate; 
    \item \revised{The slowed protoplanet's growth} suppresses the rate of expansion of the protoplanet's feeding zone, which leads to reducing the capture rate, whereas it increases the duration of protoplanet growth and thereby increases the number of conjunctions of planetesimals with the protoplanet, which leads to early capture of planetesimals.
\end{enumerate}
Our main findings are as follows:
\begin{itemize}
    \item The second effect (accelerated radial drift) always dominates over the first one (inefficient damping) in the parameter ranges considered in this study, thereby reducing the total mass of captured planetesimals relative to the case with no gap (see \refsec{sec:Model2}).
    \item The third effect (suppressed disc gas supply) makes no change or reduction in the total mass of captured planetesimals and, on the other hand, makes major planetesimal accretion occur earlier (see \refsec{sec:Model3}).
\end{itemize}
In conclusion, gap formation leads to a reduction in the total mass of captured planetesimals.

From our numerical simulations, we estimate that a Jupiter-mass planet captures, at most, 3~$\Mearth$ after the onset of runaway gas accretion until disc dispersal at 5.2~AU in the MMSN (see \refsec{sec:Discussion_Mint}). For such large amounts of heavy elements inferred for several gas giants to accrete \citep[e.g.,][]{Wahl+2017}, the solid surface density has to be more than five times higher than the MMSN value.
Also, although detailed investigation being needed, the early accretion of planetesimals that we have found in \refsec{sec:Model3} may be important for explaining the fact that heavy elements are depleted in the outer envelope and instead concentrated in the inner envelope and central core of Jupiter \citep[][]{Wahl+2017}.

For explaining dense warm Jupiters that are inferred to contain $> 100~\Mearth$ heavy elements \citep[][]{Thorngren+2016}, we would need additional capture and/or supply processes such as planetary radial migration, during which the protoplanet would encounter planetesimals outside the feeding zone \citep[][]{Alibert+2005}. 
Such an investigation will be done in our forthcoming paper.

\newpage
\appendix
\section{Benchmark Test of Our Orbital Integration Code}\label{App_BMT}
\begin{figure}
    \begin{center}
    \includegraphics[width=80mm]{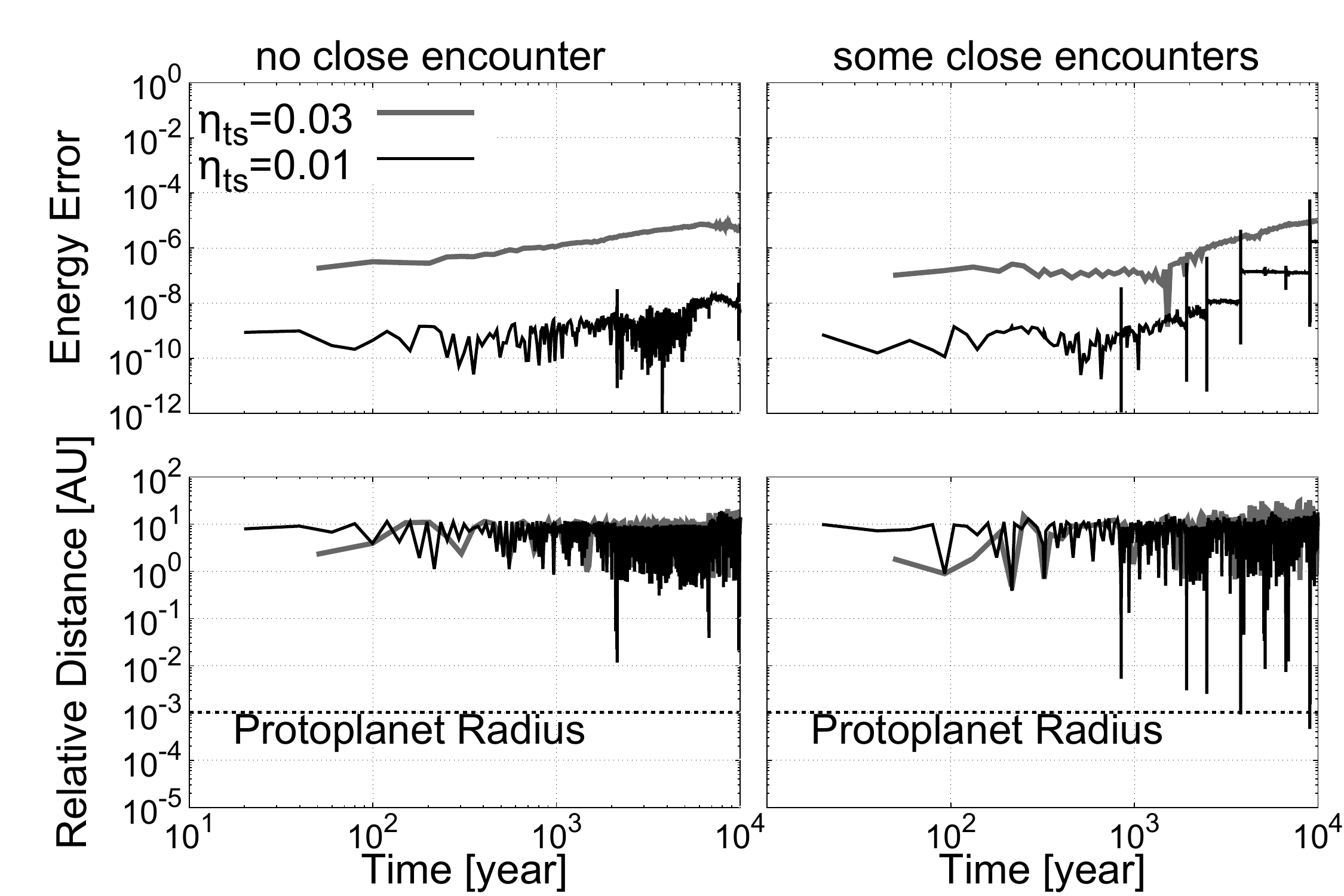}
    \caption{
    A benchmark test of our orbital integration code. 
    The upper two panels show the change in energy error defined by \refeqs{eq:App_Eng_Err}. 
    The lower two panels show the change in the relative distance between the particle and planet. 
    The planet never experiences any close encounter with the planet in the case shown by the left panels, whereas the planet does so several times in the case shown by the right panels.
    The black and grey lines show the results for $\etats=0.01$ and 0.03, respectively. 
    }
    \label{fig:App_Benchmark}
    \end{center}
\end{figure}

For a benchmark test, we consider a Jupiter-mass planet in a circular Keplerian orbit and integrate the motion of test particles that orbit a central star of solar mass.
We distribute the planetesimals using the same method of our simulation and check energy errors during numerical integration; the energy error is defined as 
\begin{equation}\label{eq:App_Eng_Err}
	\EngErr \equiv \frac{| \Ejacobi - \Ejacobizero |}{\Ejacobizero},
\end{equation}
where $\Ejacobizero$ is the initial Jacobi energy of the particle.
The energy error comes from two components, the gravitational force of the central star and that of the protoplanet.
The relative distance from the central star is almost constant and the energy error accumulates linearly with time.
On the other hand, the relative distance from the protoplanet changes largely and the energy error depends on the distance of the closest encounter.
We check the each component considering the cases of with and without close encounters with the protoplanet.
Figure \ref{fig:App_Benchmark} shows the some of the results of numerical integration.
The upper panels show the change in the energy error during numerical integration.
The lower panels show the change in the relative distance between the particle and the planet.
The left column shows the results for the case where the particle experiences no close encounter with the planet, whereas the right column shows the results for the case where the particle does so.
In the case with no close encounter, the energy error accumulates linearly with time and it is expected that the energy error would be suppressed smaller than $\sim 10^{-7}$ after the $10^5 \yrs$ integration.
On a close encounter with the planet, the energy error increases sharply and its magnitude depends on the distance. 
It turns out that increase in energy error due to close encounters is suppressed to as small as $\sim 10^{-7}$ for $\etats$ = 0.01, if $\rplpl >10^{-3} \AU$.
In our numerical study, all particles getting inside the radius of the planet are removed from the calculation and the protoplanet growth is stopped at Jupiter mass.
Thus, setting the accuracy-controlling parameter as $\etats$ = 0.01 and considering the expanding envelope which has a mean density of $0.125 \g / {\cm}^3$, the maximum energy error in the end of our simulation is as small as $\sim 10^{-7}$.

\section{Change of the Jacobi Energy}\label{App_Ejacobi}

To derive the variations of the orbital elements $\axi$, $\ecc$ and $\inc$, we consider an orbital system of cylindrical coordinates $(R, \phi, \zeta)$.
In this coordinate, the velocity of planetesimals ${\bf \vplts}$ is given as
\begin{align}\label{eq:vel_plts_co}
	\vplR &= \vKep (\axi) \frac{\ecc \sin{\tra}}{ (1-{\ecc}^2)^{1/2} }, \\
	\vplP &= \vKep (\axi) \frac{1+\ecc \cos{\tra}}{ (1-{\ecc}^2)^{1/2} }, \\
	\vplz &= 0,
\end{align}
where, $\tra$ is a true anomaly.
The velocity of disk gas ${\bf \vgas}$ is given as
\begin{align}\label{eq:vel_plts_co}
	\vgasR &= 0, \\
	\vgasP &= \vKep (\rr) \left\{1 - \etadisk (\rr) \right\} \sin \angeps, \\
	\vgasz &= -\vKep (\rr) \left\{1 - \etadisk (\rr) \right\} \cos \angeps,
\end{align}
where $\rr$ is the distance from the central axis of the circumstellar disc, and
\begin{align}\label{eq:vel_plts_co}
	\sin{\angeps} &= \frac{\cos{\inc}}{\cos{\angdel}}, \\
	\cos{\angeps} &= \sin{\inc} \frac{ \cos{(\phi + \lpc)} }{\cos{\angdel}}, \\
	\cos{\angdel} &= \frac{\rr}{r} = \frac{\rr}{\axi} \frac{ 1 + \ecc \cos{\tra}}{1 - {\ecc}^2 },
\end{align}
where $\lpc$ is a longitude of pericenter.
Using these equations, the relative velocity between the planetesimals and ambient disc gas is given as ${\bf \vplgs} = {\bf \vgas} - {\bf \vplts}$.
The variations of the orbital elements are given by
\begin{align}
	\deriv{\axi}{t} &= \frac{2}{\mm \sqrt{1-{\ecc}^2} } \left\{ \ForceR \ecc \sin \tra + \ForceP \frac{\axi}{r} \left(1 - {\ecc}^2 \right) \right\}, \\
	\deriv{\ecc}{t} &= \frac{\sqrt{ 1 - {\ecc}^2 }}{\mm \axi } \left\{ \ForceR \sin \tra + \ForceP \left(\cos \eca + \cos \tra \right) \right\}, \\
	\deriv{\inc}{t} &= \frac{1}{\mm \axi \sqrt{1-{\ecc}^2} } \Forcez \frac{r}{\axi} \cos \left(\tra + \lpc \right),
\end{align}
where ${\bf \Force}$ is the force acting on the planetesimals given by
\begin{align}
	\ForceR &= \Force \frac{\urelR}{\vplgs}, \\
	\ForceP &= \Force \frac{\urelP}{\vplgs}, \\
	\Forcez &= \Force \frac{\urelz}{\vplgs}
\end{align}
In our simulation, $\Force = \fgas$.
The averaged change of the Jacobi energy for one orbital period is given as
\begin{align}\label{eq:ave_int_dotEjacobi}
	\statave{\Ejdotnrm} &= \frac{1}{\TKep} \int_{0}^{\TKep} \Ejdotnrm {\rm d}t, \nonumber \\
	                    &= \frac{1}{2 \pi} \int_{0}^{2 \pi} \Ejdotnrm \frac{ (1- {\ecc}^2 )^{3/2} }{ \left( 1 + \ecc \cos \tra \right)^2 } {\rm d} \tra.
\end{align}
Using above equations and \refeqs{eq:der_E_Jacobi_orb}, we know the dependence of the change of Jacobi energy on the growth and damping timescales.

\section*{Acknowledgements}
We thank Tristan Guillot, Julia Venturini, Yann Alibert and Yuhiko Aoyama for valuable discussions and suggestions. 
\revised{We are also grateful to the reviewer Keiji Ohtsuki for his careful reading and constructive comments.}
This work is supported by JSPS KAKENHI Grant Numbers 17H01153 and 18H05439
and JSPS Core-to-Core Program ``International Network of Planetary Sciences''.
Numerical computations were carried out on the Cray XC50 at the Center for Computational Astrophysics, National Astronomical Observatory of Japan.


\bibliographystyle{mnras}
\bibliography{main}
\addcontentsline{toc}{section}{References}

\bsp	
\label{lastpage}
\end{document}